\documentclass[twocolumn,showpacs,superscriptaddress,amssymb]{revtex4}

\usepackage{psfrag,graphicx}
\usepackage{dcolumn}
\usepackage{bm}
\usepackage{amsfonts,amssymb,amsmath}        
\usepackage{slashed}

\usepackage{xcolor}

\newcommand{\be}{\begin{equation}}
\newcommand{\ee}{\end{equation}}
\newcommand{\bq}{\begin{eqnarray}}
\newcommand{\eq}{\end{eqnarray}}

\newcommand{\ALPHA}{\mbox{\boldmath${\alpha}$}}
\newcommand{\GAMMA}{\mbox{\boldmath${\gamma}$}}

\newcommand{\pp}[2]{\frac{\partial#1}{\partial #2}}

\newcommand{\ket}[1]{\left |#1 \right\rangle}
\newcommand{\bra}[1]{\left \langle #1 \right |}
\begin{document}

\title{$(3+1)$-dimensional topological quantum field theory from a tight-binding model of interacting spinless fermions}

\author{Mauro Cirio}
\affiliation{Centre for Engineered Quantum Systems, Department of Physics and Astronomy, Macquarie University, North Ryde, NSW 2109, Australia}
\author{Giandomenico Palumbo}
\affiliation{School of Physics and Astronomy, University of Leeds, Leeds, LS2 9JT, United Kingdom}
\author{Jiannis K. Pachos}
\affiliation{School of Physics and Astronomy, University of Leeds, Leeds, LS2 9JT, United Kingdom}
\date{\today}  


\begin{abstract}

Currently, there is much interest in discovering analytically tractable $(3+1)$-dimensional models that describe interacting fermions with emerging topological properties. Towards that end we present a three-dimensional tight-binding model of spinless interacting fermions that reproduces, in the low energy limit, a $(3+1)$-dimensional Abelian topological quantum field theory called BF model. By employing a mechanism equivalent to the Haldane's Chern insulator, we can turn the non-interacting model into a three-dimensional chiral topological insulator. We then isolate energetically one of the two Fermi points of the lattice model. In the presence of suitable fermionic interactions, the system, in the continuum limit, is equivalent to a generalised $(3+1)$-dimensional Thirring model. The low energy limit of this model is faithfully described by the BF theory. Our approach directly establishes the presence of $(2+1)$-dimensional BF theory at the boundary of the lattice and it provides a way to detect the topological order of the model through fermionic density measurements.

\end{abstract}

\maketitle
\section{ Introduction}
The interest in strongly interacting fermionic systems has recently found  new applications related to topological phases of matter. In the non-interacting case a complete classification~\cite{Ryu2, Kitaev} of standard topological insulators~\cite{Kane1} of free fermions exists. Unfortunatelly, it is not possible to straightforwardly extend these results to the interacting case. For example, it is not possible to generalise the band theory approach to topological invariants, so more flexible approaches have to be invented \cite{Gurarie11}.
The introduction of interactions in a free fermion system can  either connect different phases of matter \cite{Fidkowski11} or give access to new ones \cite{Wang222}. Examples of the latter are the two-dimensional topological Mott insulators \cite{Raghu11}, where interactions can open an insulating gap and drive the system to topological phases not accessible in the non-interacting case. 

Much progress in the study of interacting fermionic systems has already been made in 1+1 and 2+1 dimensions \cite{Fidkowski22, Lu111}. In three spatial dimensions the situation is somehow less clear, though some analysis has been already carried out \cite{Wang222,Ran11}. 
Complications arise already in the effective description, where the Chern-Simons theory \cite{WenBook} only holds in even spatial dimensions with broken time-reversal symmetry. A natural generalization of Chern-Simons theory is the  topological BF theory, which is well defined in any dimensions \cite{Blau}. In two spatial dimensions BF theories can be interpreted as double Chern-Simons theories, allowing for the description of time-reversal symmetric topological insulators~\cite{Bernevig}. BF theories have also been proposed as effective theories for describing topological insulators in any dimension~\cite{Moore,Ryu3, Palumbo2, Sodano, Maciejko}. Nevertheless, very few interacting fermionic models that give rise to BF theory are available. 

Here we make another step into the exploration of  interactions-driven phases of matter. Our starting point is a cubic lattice of spinless fermions. For particular values of the couplings and in the absence of interactions the system becomes a chiral topological insulator~\cite{Ryu1}. Our approach is similar in spirit to Haldane's Chern insulator~\cite{Haldane1}, which gives us the ability to arbitrarily tune the asymmetry in the energy spectrum of the model. This allows us to enter a regime where the dynamics, associated with one of the two Dirac fermions present in the model, is adiabatically eliminated~\cite{Palumbo1,Palumbo33}. Subsequently, we introduce interactions between the tight-binding fermions to obtain a generalization of the $(3+1)$-dimensional massive Thirring model~\cite{Thirring} with a tensorial current. By applying a series of transformations~\cite{Fradkin1} we show that our system simulates a $(3+1)$-dimensional topological massive gauge theory~\cite{CremmerScherk,Allen}. The short distance behaviour of this theory is dominated by a Maxwell term. The large distance behaviour is characterised by an Abelian BF term which is topological in nature and it gives mass to the gauge field. The connection of the fermionic tight-binding model to the BF theory allows us to directly obtain that the boundary of the lattice is described by the $(2+1)$-dimensional BF theory. Finally, we identify analytical expressions for topological invariants associated with the model and relate them to physical local fermionic observables. This method allows us  to  probe the topological properties of our three-dimensional system and provides a possible platform for simulating $(3+1)$-dimensional gauge theories in the laboratory with cold atoms~\cite{Cirac2} in optical lattices~\cite{Zoller1,LewensteinBook,AlbaPachos}.\\
This article  is organized as follows. In Section \ref{section:freeFermion} a free fermion tight binding  model is introduced. We  focus  on the  kinematic sector by analysing the  (gapless) energy spectrum, the symmetry properties, and the low energy limit  of the model. We also consider the effect of additional mass terms which open a gap in the spectrum and allow us to  show the existence of a chiral topological insulating phase. In Section \ref{section:Interactions} we leave the free fermion description by introducing 4-bodies interactions in the tight binding model. We then show that in the low energy limit the model is described by bosonic degrees of freedom  and we find  the corresponding effective theory  through a duality operation. Interestingly, the effective theory contains a purely topological term. By proposing opportune bosonization rules we give a map between observables for the effective and microscopic theory. We then  explore two features of the theory in its purely topological regime. We find that the boundary of the model is  described by a topological theory. Finally, we describe microscopic fermionic observables  which can be used to test the topological features of the model.

\section{Free Fermion Model} 
\label{section:freeFermion}
\begin{figure}[h!]
\begin{center}
\includegraphics[scale=0.38]{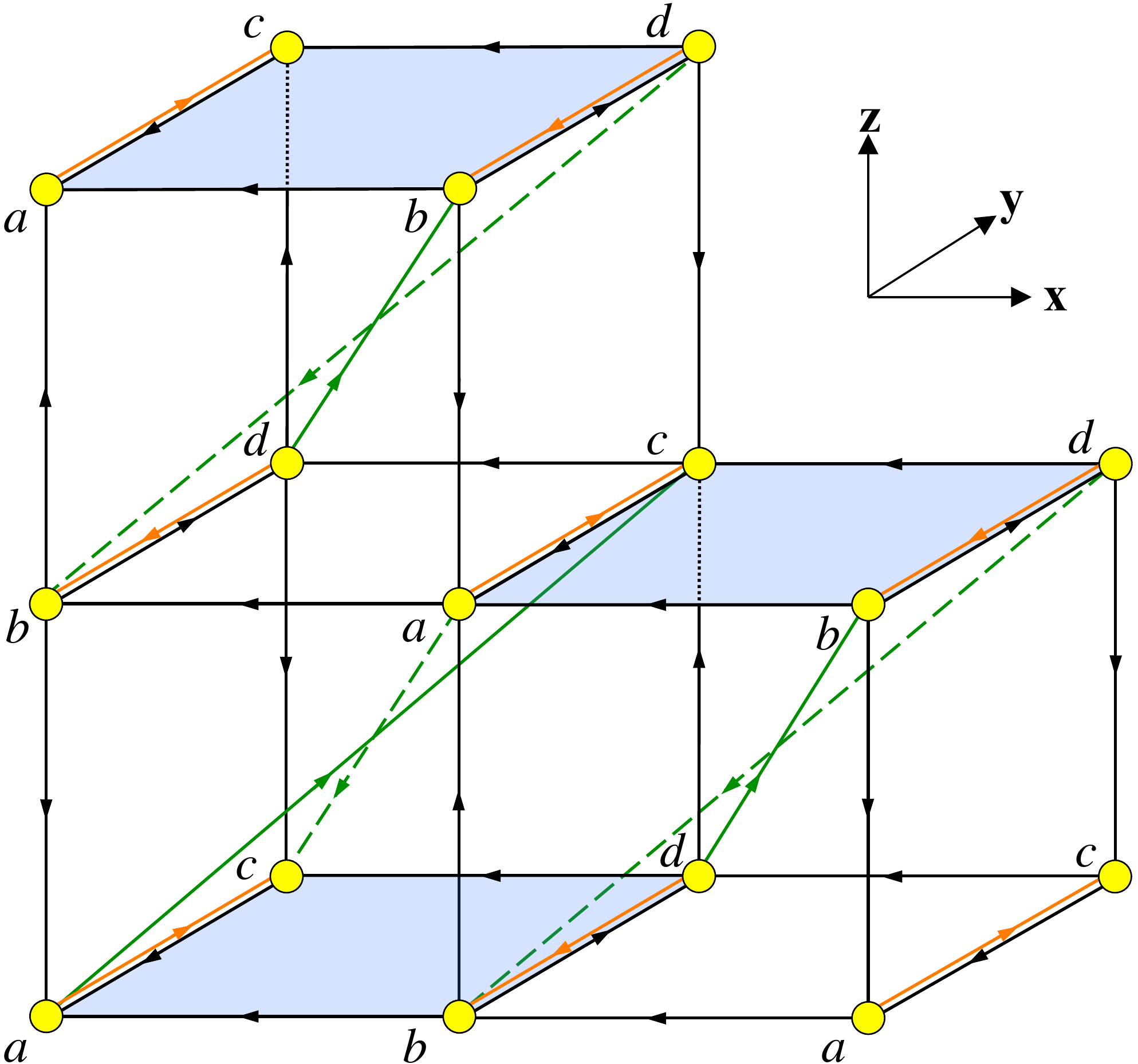}
\end{center}
\caption{\label{figCube}  The tight-binding model, where spinless fermions (yellow) reside on the vertices of a cubic lattice. The plaquette unit cell has four fermions labelled $a,b,c,d$. The fermions tunnel along the lattice via Hamiltonian (\ref{eq:hamiltonian}). Tunnelling takes place along the edges of the cubic lattice (black) with coupling $t$ and a phase that is determined by the black arrows, i.e. $\chi^t= i$ for positive and $\chi^t=- i$ for negative direction. A purely imaginary staggering term in the $y$-direction (orange) has coupling $\delta t$ and a phase $\chi^{\delta t}$ with  the same phase convention as $\chi^t$. Tunnelling along the diagonals of the cube (green) have coupling $\bar{t}$ and phase $\chi^{\bar{t}}=i \bar{t}e^{\pm i\phi}$ with $\phi\in[0,\pi/2]$, where the negative  (positive) sign is chosen for full (dashed) lines.}
\end{figure}

Let us begin with an overview of the model. We consider spinless fermions, positioned on the vertices of a three-dimensional cubic lattice $\Lambda$, as shown in Fig.~\ref{figCube}. The tight-binding Hamiltonian is given by
\begin{equation}
\label{eq:hamiltonian}
H=t\sum_{\langle {\bf i},{\bf j}\rangle}\chi^t_{{\bf ij}}f_{\bf i}^\dagger f_{\bf j}+\delta t \sum_{{\langle {\bf i},{\bf j}\rangle}_y} \chi^{\delta t}_{\bf ij} f_{\bf i}^\dagger f_{\bf j}-\frac{\bar{t}}{2}\sum_{\langle\langle\langle {\bf i},{\bf j}\rangle\rangle\rangle}\chi^{\bar{t}}_{\bf ij} f_{\bf i}^\dagger f_{\bf j},
\end{equation}
where ${\bf i},{\bf j}\in\Lambda$ and $f^\dagger_{\bf i}$  and $f_{\bf i}$ are the creation and annihilation fermion operators at position ${\bf i}$ of the lattice. We define planar unit cells populated by four fermion flavours $f\in\{a,b,c,d\}$, as shown in Fig.~\ref{figCube}. Let us analyse each term of the Hamiltonian. The first term, which we call kinematic, has coupling $t$ and corresponds to nearest-neighbour $\langle {\bf i},{\bf j}\rangle$ hopping. The phases $\chi^t$ are such to create a net $\pi$ flux through each plaquette. The term proportional to $\delta t$ describes a staggering between sites along the $y$-direction indicated by $\langle {\bf i},{\bf j}\rangle_y$.  The last term corresponds to tunnelling between the next-next-nearest neighbouring sites, $\langle\langle\langle {\bf i},{\bf j}\rangle\rangle\rangle$, with coupling $\bar{t}$. The phase factors $\chi^{t}$, $\chi^{\delta t}$ and $\chi^{\bar{t}}$ are defined in Fig.~\ref{figCube}. Let us now study this model more explicitely.\\
The lattice of the unit cells (in blue in Fig. \ref{figCube}) is given by:  $\bar{\Lambda}=\{{\bf i}\in\mathbb{R}: {\bf i}=n_1 {\bf s}_1+n_2 {\bf s}_2+n_3 {\bf s}_3\}$, with  $n_i\in\mathbb{N}$  and ${\bf s}_1=(2,0,0) $, ${\bf s}_2=(0,2,0)$, ${\bf s}_3=(1,0,1)$ written in units of a fixed reference length.
The Hamiltonian in Eq. (\ref{eq:hamiltonian}) can be written as
\begin{equation}
\label{eq:hamiltonian2}
H= H_0+H_m\;\;,
\end{equation}
where $H_0=t\sum_{\langle {\bf i},{\bf j}\rangle}\chi^t_{{\bf ij}}f_{\bf i}^\dagger f_{\bf j}$ is a kinematic Hamiltonian (defined through the black links in Fig. \ref{figCube}) which has gapless spectrum. In order to open a gap in the model we introduce  $H_m=\delta t \sum_{{\langle {\bf i},{\bf j}\rangle}_y} \chi^{\delta t}_{\bf ij} f_{\bf i}^\dagger f_{\bf j}-\frac{\bar{t}}{2}\sum_{\langle\langle\langle {\bf i},{\bf j}\rangle\rangle\rangle}\chi^{\bar{t}}_{\bf ij} f_{\bf i}^\dagger f_{\bf j}$ which is defined along the red and green links in Fig. \ref{figCube}. Let us now  define the two terms of the Hamiltonian one by one.
\subsection{The kinematic model}
\label{section:kinematicModel}
As can be seen by inspecting Eq. (\ref{eq:hamiltonian2}) and Fig. \ref{figCube}, the kinematic Hamiltonian of the model can be written as
\begin{equation}
\label{eq:KineticHamiltonian2}
\begin{array}{lll}
H_0&=&it\sum_{{\bf i}}\left[(-a^\dagger_{{\bf i}} b_{{\bf i}}+b^\dagger_{{\bf i}} d_{{\bf i}}+d^\dagger_{{\bf i}}c_{{\bf i}}+c^\dagger_{{\bf i}}a_{{\bf i}})\right.\\
&&+(a^\dagger_{{\bf i}+{\bf{s}}_1}b_{{\bf i}}+ c^\dagger_{{\bf i}+{\bf{s}}_1}d_{{\bf i}}+d^\dagger_{{\bf i}}b_{{\bf i}+{\bf{s}}_2}+a^\dagger_{{\bf i}+{\bf{s}}_2}c_{{\bf i}})\\
&&\left.+(b^\dagger_{{\bf i}+{\bf{s}}_3-{\bf{s}}_1}a_{{\bf i}}+b^\dagger_{{\bf i}} a_{{\bf i}+{\bf{s}}_3}+d^\dagger c_{{\bf i}+{\bf{s}}_3}+d^\dagger_{{\bf i}+{\bf{s}}_3-\bold{s}_1}c_{{\bf i}})\right]\\
&&+\text{h.c.}\;\;,
\end{array}
\end{equation}
where ${\bf i}\in\bar{\Lambda}$ is intended and where $t$ is an energy scale. This Hamiltonian  is known \cite{Susskind1} to give rise in the continuum limit to two massless Dirac fermions. Let us now calculate the spectrum explicitely.\\
The reciprocal lattice is defined as $\bar{\Lambda}_p=\{{\bf p}\in\mathbb{R}^3:{\bf p}=\sum_i n_i {\bf p}_i\}$ where the vectors ${\bf p}_i$ satisfy ${\bf p}_i \cdot{\bf s}_j=2\pi\delta_{ij}$ and are explicitely defined to be ${\bf p}_1=\pi(1,0,-1)$, ${\bf p}_2=\pi(0,1,0)$, ${\bf p}_3=2\pi(0,0,1)$.
The Brillouin zone ($\text{BZ}$) is defined as the elementary cell in the reciprocal lattice  $\text{BZ}=\{{\bf p}\in\bar{\Lambda}_p: {\bf p}=p_i{\bf p}_i\}$ with $p_i\in[0,1)$. A generic vector in the Brillouin zone can be written as ${\bf p}\equiv(p_x,p_y,p_z)=\pi(p_1,p_2,2p_3-p_1)$. The periodic invariance of the phase space allows us to parametrize the Brillouin zone in a different and somehow more convenient way. We can in fact define it as $\text{BZ}=\{{\bf p}\in\bar{\Lambda}_p: {\bf p}=(p_x,p_y,p_z)$ with $p_x\in[0,\pi)$, $p_y\in[0,\pi)$, $p_z\in[0,2\pi)$ where the volume of the Brillouin zone is  $2\pi^3$. We now have all the ingredients to define the Fourier transform $a_{{\bf r}}=\sum_{{\bf p}\in\text{BZ}}e^{-i{\bf p}\cdot {\bf r}}a_{{\bf p}}$ and analogously for $b,c,d$.
By introducing the Fourier transformed operators in Eq. (\ref{eq:KineticHamiltonian2}) we find
\begin{equation}
\begin{array}{lll}
H_0&=&it\sum_{{\bf p}}\left[(-1+e^{2 i p_x}-e^{-i (p_z-p_x)}-e^{ i (p_z+p_x)})a^\dagger_{{\bf p}} b_{{\bf p}}\right.\\
&&+(e^{2 i p_y}-1)a^\dagger_{{\bf p}} c_{{\bf p}} + i (1-e^{2 i p_y})b^\dagger_{{\bf p}} d_{{\bf p}}\\
&&+\left.(-1+e^{2 i p_x}-e^{ i (p_z+p_x)}-e^{- i (p_z-p_x)})c^\dagger_{{\bf p}} d_{{\bf p}}\right]\\
&&+\text{h.c.}\\
&=&\sum_{{\bf p}}{\Psi^\prime}^\dagger\bar{H}^\prime_0\Psi^\prime\;\;,
\end{array}
\end{equation}
with
\begin{equation}
\Psi^\prime=\left(\begin{array}{c}a_{{\bf p}}\\b_{{\bf p}}\\c_{{\bf p}}\\d_{{\bf p}}\end{array}\right)\;\;,
\end{equation}
 and with the kernel $\bar{H}_0^\prime$ given by
 \begin{equation} 
 \label{eq:BAndC}
 \bar{H}^\prime_0=t\left(\begin{array}{cccc}0&B&C&0\\B^*&0&0&-C\\C^*&0&0&B\\0&-C^*&B^*&0\end{array}\right)\;\;,
 \end{equation}
 where we have defined $B=i(-1+e^{2ip_x}-e^{-i(p_z-p_x)}-e^{i(p_z+p_x)})$ and $C=i(e^{2ip_y}-1)$.
 From the explicit expression of  $H_0$ and from Fig. \ref{figCube} we can easily see that the set of vertices $a$ and $d$ only interacts with the set $b$ and $c$. This condition defines a chiral symmetry. In fact, such a symmetry describes the existence of a bipartition of the lattice ``broken'' by \emph{all} couplings (see Appendix \ref{AppDiscreteSymmetries} for more details). The existence of chiral symmetry allows to cast the Hamiltonian in an off-block diagonal form. In our case this is easily seen: after the definition of a new basis
 \begin{equation}
 \label{eq:ChiralBasis}
\Psi=\left(\begin{array}{c}a_{{\bf p}}\\d_{{\bf p}}\\c_{{\bf p}}\\b_{{\bf p}}\end{array}\right)\;\;,
\end{equation}
the Hamiltonian takes the form
\begin{equation}
H_0=\Psi^\dagger\bar{H}_0\Psi\;\;,
\end{equation}
with
  \begin{equation}
 \label{Kernelb0} \bar{H}_0=t\left(\begin{array}{cccc}0&0&C&B\\0&0&B^*&-C^*\\C^*&B&0&0\\B^*&-C&0&0\end{array}\right)\;\;.
 \end{equation}
 This Hamiltonian has eigenvalues (with degeneracy 2) given by
 \begin{equation}
 \label{eq:SpectrumHKinetic}
 E_0=\pm t\sqrt{6-2\cos{2p_x}-2\cos{2p_y}+2\cos{2p_z}}\;\;.
 \end{equation}
The spectrum has then two double degenerate bands and it becomes gapless at two Fermi points where the two bands touch each other.
 The  two independent Fermi points are  given by
 \begin{equation}
 \label{eq:FermiPoints}
 \left\{\begin{array}{lll}
 {\bf P_+}&=&(0,0,\frac{\pi}{2})\\
{ \bf P_-}&=&(0,0,\frac{\pi}{2}+\pi)\;\;.
 \end{array}\right.
 \end{equation}
In order to study the behaviour around the Fermi points we now define the following matrices
  \begin{equation}
  \begin{array}{c}
  \label{eq:alphaM2}
 \begin{array}{cc}
\alpha^x= \left(\begin{array}{cccc}0&0&0&-1\\0&0&-1&0\\0&-1&0&0\\-1&0&0&0\end{array}\right)\;,&\alpha^y= \left(\begin{array}{cccc}0&0&-1&0\\0&0&0&1\\-1&0&0&0\\0&1&0&0\end{array}\right)\;,
  \end{array}\\
  \\
   \begin{array}{c}
 \alpha^z=\left(\begin{array}{cccc}0&0&0&+i\\0&0&-i&0\\0&i&0&0\\-i&0&0&0\end{array}\right)\;\;,
  \end{array}
  \end{array}
 \end{equation}
 which  satisfy the algebra
  \begin{equation}
  \begin{array}{lll}
 \{\alpha^i,\alpha^j\}&=&2\delta^{ij}
 \end{array}\;\;.
 \end{equation}
 We now  introduce coordinates around the Fermi points $\hbar{\bf p}=\hbar {\bf P_{\pm}}+(k_x,k_y,k_z)$ for small $k_x$, $k_y$ and $k_z$, so that the Hamiltonian around the Fermi points looks like
 \begin{equation}
 \label{eq:DiracEquation2}
 \bar{\mathcal{H}}^0_\pm=c(k_x \alpha_x+k_y \alpha_y\pm k_z \alpha_z)\;\;,
 \end{equation}
 where $c=2 t  /\hbar$. The Hamiltonians in Eq. (\ref{eq:DiracEquation2}) represent two massless Dirac fermions. 
 \subsubsection{Symmetries}
 \label{section:symmetries}
The symmetries of the kinematic model can be studied by analyzing the Hamiltonian kernel (\ref{Kernelb0}). In particular we are interested in 
 checking the behaviour of the model under time-reversal, particle-hole and chiral symmetry. For an introduction to the defintions of these symmetries we refer to Appendix \ref{AppDiscreteSymmetries}. In the table below we express the conditions on the Hamiltonian kernel under which these symemtries are satisfied.
 
 \begin{table}[h!]
\centering
\begin{tabular}{cc}
Symmetry&Condition\\
\hline
&\\
Time-Reversal&$\bar{H}({\bf p})=\bar{H}^{*}(-{\bf p})$\\
Particle-Hole &$\bar{H}({\bf p})=-\bar{H}^{*}(-{\bf p})$ \\
Chiral & $\exists \bar{C}_s : \bar{C}_s^\dagger=\bar{C}_s^{-1}: \bar{C}_s \bar{H}({\bf p})\bar{C}^\dagger_s=-\bar{H}({\bf p})$\\
\hline
\end{tabular}
\label{ciao}
\end{table}

\noindent  Inspection of  the Hamiltonian kernel given in  (\ref{Kernelb0}) shows us that $\bar{H}_0^*(-{\bf p})=-\bar{H}_0({\bf p})$. This condition means that the system breaks time-reversal symmetry and preserves particle-hole symmetry. We also have an explicit chiral symmetry since the Hamiltonian anticommutes with the matrix $\bar{C}_s$ defined as
 \begin{equation}
 \bar{C}_s=\left(\begin{array}{cccc}1&0&0&0\\0&1&0&0\\0&0&-1&0\\0&0&0&-1\end{array}\right)\;\;,
 \end{equation}
 which is hermitian and unitary, as expected from the block structure of  Hamiltonian   (\ref{Kernelb0}).
 \subsection{Gapped model}
 \label{section:gappedModel}
   \begin{figure}
\begin{center}
\includegraphics[scale=0.5]{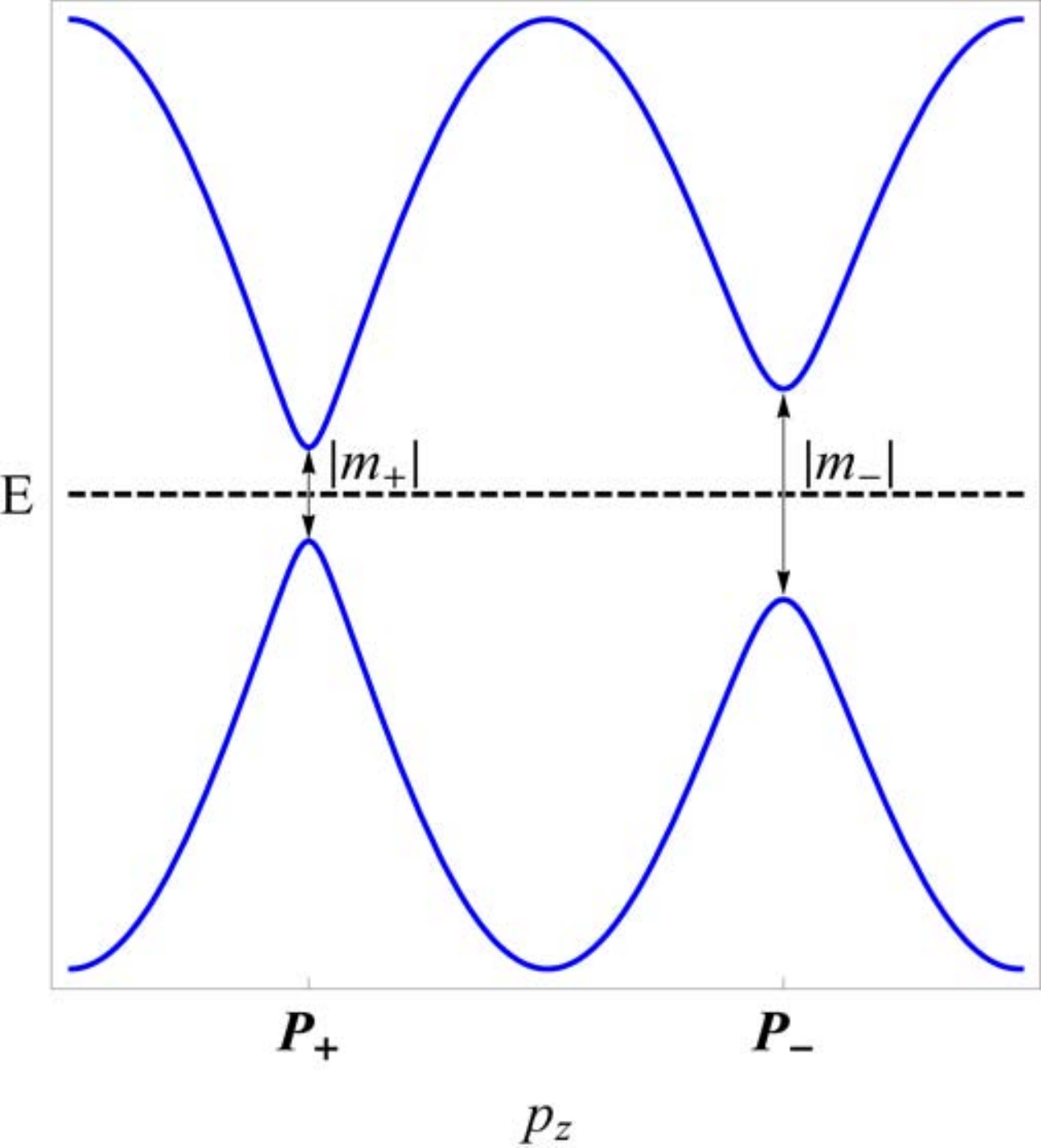}
\end{center}
\caption{\label{fig:KinematicBandsMass} Energy bands for the full model described in Eq. (\ref{eq:hamiltonian}) as a function of the momentum variable $p_z$ for fixed $p_x=p_y=0$, in arbitrary units. The parameters $\delta t$ and $\bar{t}$ are tuned to open a gap. The model is in fact a description of a chiral topological insulator. The dashed line represents the Fermi energy and  highlights the insulating properties of the material.
}
\end{figure}
 The kinematic model introduced  in the previous section is gapless.We now introduce a gap term. In this way the low energy physics of the model is described  by a massive  Dirac fermion. Such a  mass term  has to anticommute with all the $\alpha$ matrices (Eq. (\ref{eq:alphaM2})),  square to the identity and we also require it to satisfy  chiral symmetry. As can be  checked, the mass term has to be proportional to $\beta=\bar{C}_s\alpha^x\alpha^y\alpha^z$.
In the chosen representation, we have
   \begin{equation}
   \label{eq:betaAndm5}
 \begin{array}{c}
 \beta= \left(\begin{array}{cccc}0&0&i&0\\0&0&0&i\\-i&0&0&0\\0&-i&0&0\end{array}\right)\;\;.
  \end{array}
 \end{equation}
The  implementation of such a mass term requires the introduction of additional couplings between the sites $a$ and $c$ and between $b$ and $d$ (as can be seen by inspecting the explicit form of $\beta$ in the basis given by Eq. (\ref{eq:ChiralBasis})). We introduce a staggering of the $a,c$ and $b,d$ couplings along the $y$ axis and a next-next nearest neighbor (NNN) interactions as shown in Fig. \ref{figCube}. The staggering NNN interactions give an equal (opposite) mass term to the two Dirac fermions defined in Eq. (\ref{eq:InteractionHamiltonian2}). Explicitely we define
 \begin{equation}
 \label{eq:InteractionHamiltonian2}
 \begin{array}{lll}
 H_m&=&\sum_{{\bf{r}}}i\delta t a^\dagger_{{\bf{r}}} c_{{\bf{r}}}+i\delta t d^\dagger_{{\bf{r}}} b_{{\bf{r}}}\\
 &&+\frac{\bar{t}}{2}\sum_{{\bf{r}}} i e^{-i \phi} a^\dagger_{{\bf{r}}} c_{{\bf{r}}+{\bf{s}}_3}+i e^{i \phi} a^\dagger_{{\bf{r}}} c_{{\bf{r}}-{\bf{s}}_3}\\
 &&+\frac{\bar{t}}{2}\sum_{{\bf{r}}}i e^{-i \phi} d^\dagger_{{\bf{r}}} b_{{\bf{r}}+{\bf{s}}_3}+i e^{i \phi} d^\dagger_{{\bf{r}}} b_{{\bf{r}}-{\bf{s}}_3}\\
 &&+\text{h.c.}\;\;,
 \end{array}
 \end{equation}
 where we have introduced different energy scales $\delta t$ and $\bar{t}$ for the staggering and NNN term respectively. We also introduced a phase $\phi$ associated with the NNN couplings. In momentum space this Hamiltonian becomes
  \begin{equation}
 \begin{array}{lll}
 H_m&=&\sum_{{\bf p}}i(\delta t+ \bar{t} \cos{(p_z+p_x+\phi)}) a^\dagger_{{\bf p}} c_{{\bf p}}\\
 &&+i(\delta t+ \bar{t} \cos{(p_z+p_x+\phi)}) d^\dagger_{{\bf{p}}} b_{{\bf{p}}}+\text{h.c.} \;\;.\end{array}
 \end{equation}
 In the basis of Eq. (\ref{eq:ChiralBasis}) the kernel in momentum space of the interaction Hamiltonian $H_1$ reads
 \begin{equation}
 \label{Kernelb1}
\bar{H}_m=(\delta t+ \bar{t} \cos{(p_z+p_x+\phi)})\beta\;\;.
 \end{equation}
This term  is  proportional to the matrix $\beta$ so it can be interpreted as a fermion mass as discussed above.\\
Now, we first notice that, when $\bar{t}\neq 0$ and $\phi\neq 0,\pi$ the particle-hole symmetry is broken:  $\bar{H}_m^*(-{\bf k})\neq-\bar{H}_m({\bf k})$.  Incidentally, it is important to notice that by adding these interactions we did not restore time-reversal symmetry (already broken in the kinematic model) in the full model. \\
The results of this section  imply that the full Hamiltonian $H=H_0+H_m$ breaks time-reversal and particle-hole symmetry while it is symmetric under chiral symmetry. We also note that the joint presence of staggering and NNN interactions allows to arbitrarly tune the fermion masses at the two Fermi points. This is easily seen by evaluating Eq. (\ref{Kernelb1}) at the two Fermi points to get two independent masses. More precisely, let us define
 \begin{equation}
 \label{eq:defMasses}
\left\{ \begin{array}{lll}
 m_+ c^2&=&\delta t+\bar{t}\cos{(\frac{\pi}{2}+\phi)}\\
 m_- c^2&=&\delta t+\bar{t}\cos{(\frac{3\pi}{2}+\phi)}\;\;,
 \end{array}\right.
 \end{equation}
 which, for the choice $\phi=\frac{\pi}{2}$ becomes
  \begin{equation}
  \label{eq:massesFullModel}
\left\{ \begin{array}{lll}
 m_+ c^2&=&\delta t-\bar{t}\\
 m_- c^2&=&\delta t+\bar{t}\;\;.
 \end{array}\right.
 \end{equation}
With these definitions we get the expression for the full Hamiltonian around the two Fermi points (to be compared with Eq. (\ref{eq:DiracEquation2}))
 \begin{equation}
 \label{eq:DiracEquationMass2}
 \mathcal{H}_\pm({\bf k})=\bar{\Psi}_\pm(c {\ALPHA}\cdot{\bf k}+m_\pm c^2 \beta )\Psi_\pm\;\;,
 \end{equation}
 where $\ALPHA=\{\alpha_x,\alpha_y,\alpha_z\}$ and ${\bf k}=\{k_x,k_y,k_z\}$.
 Notice that when $\delta t =0$ or $\phi=0,\pi$ such an arbitrary tuning would not be possible and we would get $m_+=m_-$.\\
We end up this section with the book-keeping explicit expression for the total Hamiltonian of the model of Eq. (\ref{eq:hamiltonian}). From Eq. (\ref{Kernelb0}) and (\ref{Kernelb1}) and with the definitions  (\ref{eq:alphaM2}), (\ref{eq:betaAndm5}) and the ones below Eq. (\ref{eq:BAndC}) the kernel $\bar{H}$ of the total Hamiltonian reads
 \begin{equation}
 \label{eq:FullHamiltonianFull}
 \begin{array}{lll}
 \bar{H}&=&\bar{H}_0+\bar{H}_m\\
 &=&t(\sin{2 p_x}-\sin{(p_x+p_z)}-\sin{(p_x-p_z)})\alpha^x\\
 &&+t\sin{2p_y}\alpha^y\\
 &&+t(\cos{2p_x}-\cos{(p_x+p_z)}-\cos{(p_x-p_z)}-1)\alpha^z\\
 &&+(t\cos{2p_y}-t+\delta t+ \bar{t} \cos{(p_x+p_z+\phi)})\beta\;\;.
 \end{array}
 \end{equation}
 The spectrum of the total Hamiltonian has two double degenerate bands
 \begin{equation}
 E=\pm\sqrt{(4 - 2 \cos2 p_x+ 2 \cos2 p_z + |M|^2)}\;\;,
 \end{equation}
 where $M= \left(e^{2 i p_y}-1+\delta t + t \cos{(p_x+p_y+\phi)}\right)$.
 \subsection{Chiral Topological Insulator}
\label{WindingNumber}
 Symmetry protected phases of matter for models described by a free fermion model are completely classified \cite{Ryu2}. This classification characterizes phases of matter within 10 different symmetry classes determined by the symmetry properties under time-reversal, particle-hole, and chiral symmetry. 
More specifically, one starts by continuously deforming  the Hamiltonian $H$ that describes a free fermion model to a ``reference'' Hamiltonian  $Q$ \cite{RyuArxiv}. This Hamiltonian has all occupied (empty) bands ``flatten'' with energy $+1$ (-1) in the whole Brillouin zone. This can be done by defining the operator $Q(k)$ as
\begin{equation}
Q(k)=2 P(k)-\mathbb{I}_{n+m}\;\;,
\end{equation}
with
\begin{equation}
P(k)=\sum_{i=1}^m\ket{u_i(k)}\bra{u_i(k)}\;\;,
\end{equation}
where $u_i(k)$ $i=1,\dots,m$ are the eigenvalues of the occupied bands for the total Hamiltonian and $m(n)$ is the number of occupied (empty) bands. The operator $Q$ is such that $Q^\dagger=Q$, $Q^2=\mathbb{I}$ and $\text{tr}(Q)=m-n$. This operator has eigenvalues $+1$ and $-1$ corresponding to occupied and empty bands. Each of the 10 symmetry classes mentioned above determines a manifold $\mathcal{B}$ such that $Q:\text{BZ}\rightarrow \mathcal{B}$. Within each symmetry class (and hence for each manifold $\mathcal{B}$), we  want to classify the phases of matter described  by the reference Hamiltonians $Q$. Two Hamiltonians belong to the same phase if they can be continuously deformed one into the other without encountering  a critical point. As shown in \cite{Ryu2} one can classify such phases through the $d-$th homotopy group $\pi_d$ of the manifold $\mathcal{B}$ where $d$ is the spatial dimension of the model. For example, for the symmetry class A (all symmetries broken), we have that $\mathcal{B}$ is isomorphic to the Grassmannian: $\mathcal{B}\simeq  G_{n,n+m}(\mathbb{C})\equiv U(n+m)/(U(n)\times U(m))$. In fact, the collection of all energy eigenvectors describes an element of $U(n+m)$ modulo  the ``gauge'' symmetry relabeling the eigenvectors corresponding to  occupied and empty bands. Now, for two spatial dimensions we have an infinite number of different phases as implied by $\pi_2(G_{n,n+m}(\mathbb{C}))=\mathbb{Z}$ (specifying, for example, the number of edge states for the quantum Hall effect, which in fact, being a Chern insulator,  belongs to the symmetry class A). In three spatial dimensions we have $\pi_3(G_{n,n+m}(\mathbb{C}))=e$ (where $e$ represent the group trivial element) so that only the trivial phase is allowed. Such models can become non-trivial when more symmetries are considered.  Specifically,  we are interested in the symmetry class AIII where only chiral symmetry is preserved. in this case $n=m$ (positive and energy eigenstates come in pairs, see Appendix \ref{AppDiscreteSymmetries}),  and one can write $Q$ in the following block form \cite{Ryu2}
\begin{equation}
\label{eq:Q}
Q({\bf p})=\left(\begin{array}{cc}0&q({\bf p})\\q^\dagger({\bf p}) &0\end{array}\right)\;\;,
\end{equation}
\begin{figure}[h!]
\begin{center}
\includegraphics[scale=0.6]{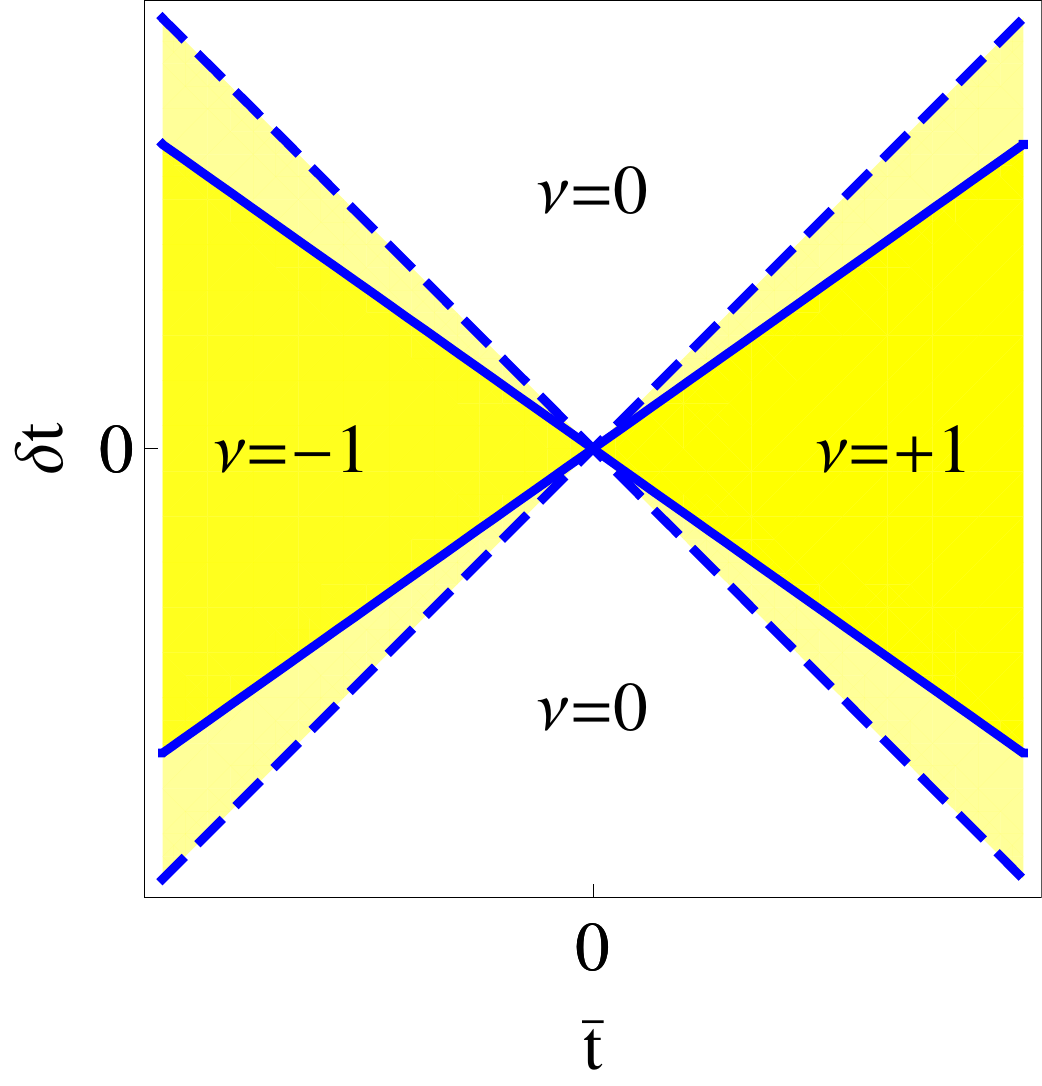}
\end{center}
\caption{\label{figCartoon} Phase diagram of Hamiltonian (\ref{eq:hamiltonian}) as a function of the couplings $\delta t$ and $\bar{t}$ parameterised by the phase $\phi$. The gapped regions with non-trivial winding number $\nu=\pm 1$ (yellow) are separated by phase transition (blue lines) from the topologically trivial regions with $\nu=0$ (white). The winding number, $\nu$, is correlated with the sign of $m_+\cdot m_-$, in the same way as in Haldane's model. The critical line for $\phi=\pi/2$ (dashed blue) and for generic value of $\phi\in[0,\pi/2]$ (solid blue) are depicted. }
\end{figure}
with $q({\bf p})\in U(n)$. Our model is then described by the function $q:\text{BZ}\rightarrow \mathcal{B}$ where $\mathcal{B}\simeq U(n)$. Contrary to the class A example, we now find that  $\pi_3(U(n))=\mathbb{Z}$ allowing for non-trivial phases in three spatial dimensions. In fact, we can define \cite{Ryu2}  a winding number $\nu$ (associated with the map $q$)  labelling all the possible phases as
\begin{equation}
\label{eq:nu}
\nu=\frac{1}{24 \pi^3}\int d^3k \epsilon^{abc}\text{tr}[(q^{-1}\partial_a q)(q^{-1}\partial_b q)(q^{-1}\partial_c q)]\;\;,
\end{equation}
where $a,b,c=1,2,3$.
\subsubsection{Phase Diagram}
We now want to see under which conditions on the parameters of the full Hamiltonian of our model (Eq. (\ref{eq:hamiltonian})) we can get non-zero winding number (Eq. (\ref{eq:nu})).\\
 Let us first resume what we learned so far. First of all, from the results of section \ref{section:gappedModel} we know that non-trivial topological properties are forbidden for $\delta t = 0$ or $\phi=\{0,\pi\}$.   In fact, in such a regime,  the particle-hole symmetry is not broken leading to the impossibility to define a winding number as shown in \cite{Ryu2}. Second,  we know that the conditions $\delta t= \bar{t}\cos{(\frac{\pi}{2}+\phi)}$ and $\delta t= \bar{t}\cos{(\frac{3\pi}{2}+\phi)}$ imply (see Eq. (\ref{eq:defMasses})) that either $m_+=0$ or $m_-=0$ meaning that the system is critical. 
Then, in these cases we expect  the winding number in Eq. (\ref{eq:nu}) to be  not  well defined.\\
%
Let us now draw the phase diagram of the model in the parameter space $\{\delta t ,\bar{t}\}$ (see Fig. \ref{figCartoon}). 
A gapless system is described 
by  imposing the equations $m_+=0$ and $m_-=0$ (see  Eq. (\ref{eq:defMasses})). Pictorially, these equations are two straight lines in the $\delta t,\bar{t}$ plane, parametrized by the phase $\phi$. They divide the parameter space $\delta t, \bar{t}$ in four  disconnected regions. We studied the behaviour of the winding number in these four regions and we found that two of them are in fact non-trivial with $\nu=\pm 1$ (see Fig. \ref{figCartoon}). When the parameter $\phi$ tends to $0$ the two non-trivial phases disappear as the two critical lines merge together. This result is consistent to the fact that $\phi=0$ corresponds to a system where particle-hole symmetry is not broken (see the analysis following Eq. (\ref{Kernelb1})) which is a sufficient condition for the absence of topological order \cite{Ryu2}. 
The non-triviality of the winding number Eq. (\ref{eq:nu}) (for a certain parameters regime) shows that the system is a  chiral  topological insulator.\\
To sumarize, the introduction of NNN neighbour interactions ($\bar{t}\neq  0$) breaks particle-hole symmetry (provided that $\phi\neq 0$) and gives an opposite contribution to the masses in Eq. (\ref{eq:defMasses}). On the contrary, the staggered interactions ($\delta t\neq 0$) give  an equal contribution to the fermion masses. As a consequence, the simultaneous presence of both interactions allow to arbitrarily tune the masses $m_\pm$. \\
 Intuitively, this model presents several formal analogies with the Haldane model \cite{Haldane1} where spinless electrons  hop on the verteces of a honeycomb lattice. Such a model has  a kinematic term which preserves time-reversal and  inversion symmetries (and breaks particle-hole symmetry) and that  gives rise to two gapless Fermi points. In addition,  next-nearest neighbour interactions (mimicking a nested magnetic field) and a staggered chemical potential break, rispectively, time-reversal and inversion symmetry. The breaking of each symmetry allows for a non-zero energy gap to appear. More precisely, the fermions at the two Fermi points acquire the same (opposite) mass due to the breaking of time-reversal (inversion) symmetry. In this sense, our NNN neighbour and staggering terms mimic, respectively,  the staggered magnetic field and the chemical potential of the Haldane model. In the light of the classification given in \cite{Ryu2}, the key-feature to build a non-trivial topological phase on top of our (Haldane) kinematic theory is to break the  particle-hole (time-reversal) symmetry. Despite these similarities, the Haldane model breaks all symmetries (it describes the physics of the quantum Hall effect without magnetic field) while we have to pay extra attention to preserve chiral symmetry in order to protect the topological phase in $3+1$ dimensions.\\
\section{Interacting Fermions Model}
\label{section:Interactions}
\begin{figure}
\begin{center}
\includegraphics[scale=0.5]{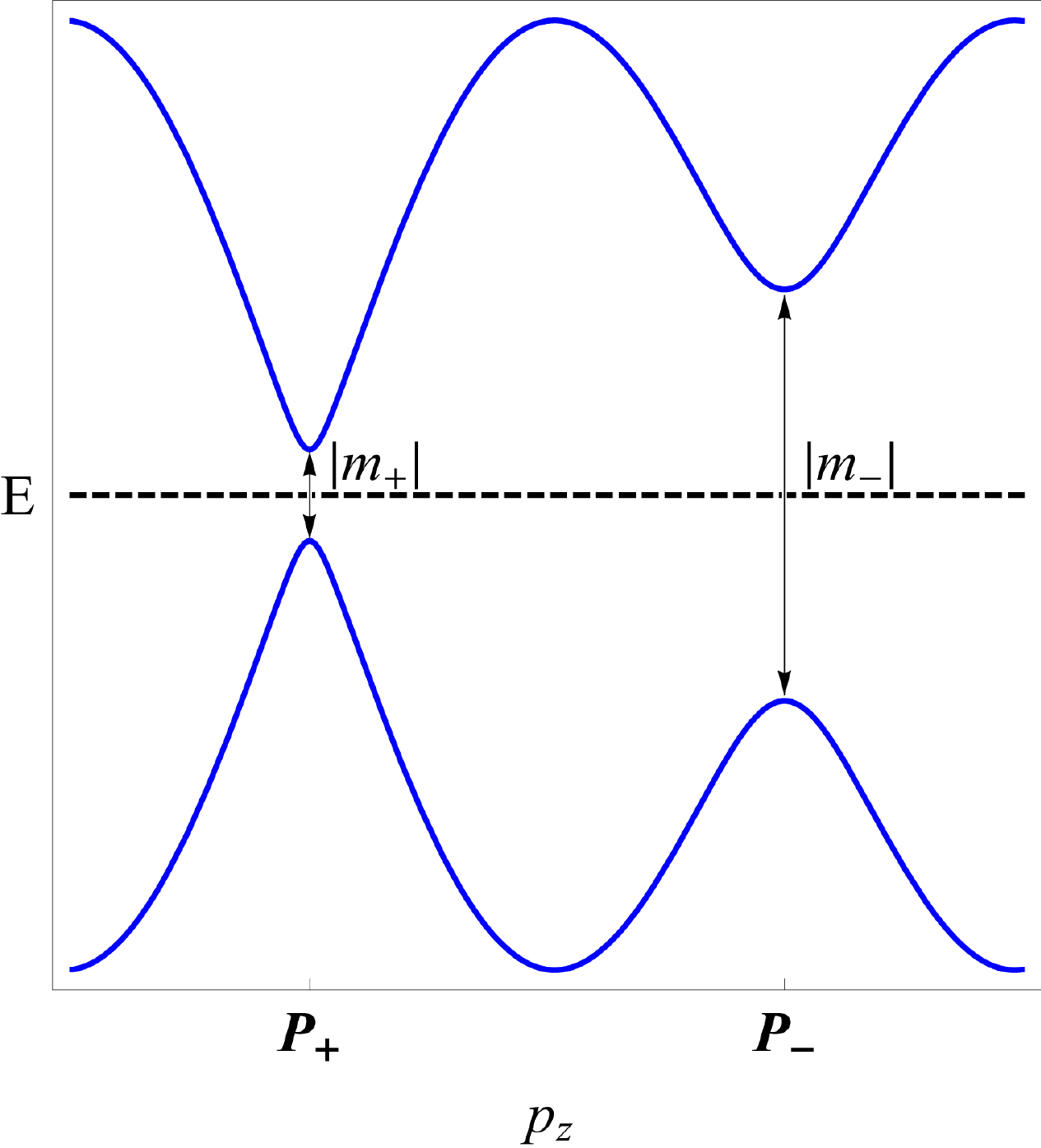}
\end{center}
\caption{\label{figAdiabaticElimination} Adiabatic elimination of a Fermi point. The energy spectrum (in arbitrary units) of the model described by the Hamiltonian in Eq. (\ref{eq:hamiltonian}) (as a function of the momentum variable $p_z$ for fixed $p_x=p_y=0$) allows  to  arbitrary tune the masses around the two Fermi points (see Eq. (\ref{eq:defMasses})). In particular, by opportunely choosing the parameters $\delta t$ and $\bar{t}$, it is possible to work in a regime where $m_-\gg m_+$. If we suppose not to have any perturbation with energy scale bigger than $|m_-|, $ then the low energy physics of the system is completely described around the $P_+$ Fermi point in the Brillouin zone.}
\end{figure}

We now turn to the case of interacting fermions. The starting point is the effective theory described in Eq. (\ref{eq:DiracEquationMass2}) with $\phi=\pi/2$. This model has enough flexibility to allow us to arbitrarily tune the masses around the two Fermi points shown in Fig.~\ref{figAdiabaticElimination}. Following the approach in \cite{Palumbo1} we can define a hierarchy in the energy scales, given by $|m_+ c^2| \ll |m_- c^2|$, and adiabatically eliminate the physics around the second Fermi point, ${\bf P}_-$. We now introduce four-body fermionic interactions with coupling $U$ that is small compared to the energy scale of ${\bf P}_-$, i.e. $\sqrt{(\hbar c)^3/U}\ll m_- c^2$, and comparable to $|m_+ c^2|$. These interactions are particularly designed so that they give rise to self-interacting current-current terms in the single Dirac fermion description corresponding to ${\bf P}_+$. The resulting effective physics is encoded in the Hamiltonian
\begin{equation}
\label{eq:Thirring}
\mathcal{H}({\bf p})=\Psi^\dagger(c {\ALPHA}\cdot{\bf p}+m c^2 \beta)\Psi+\frac{g^2}{2 m}(2J_{\mu\nu}J^{\mu\nu}-J_\mu J^\mu),
\end{equation}
where  $m\equiv m_+=(\delta t-\bar{t})/c^2$ and $g^2=2 m U$. There are two types of currents given by $J^\mu=\bar{\Psi}\gamma^\mu\Psi$ and $J^{\mu\nu}=\bar{\Psi}\gamma_5[\gamma^\mu,\gamma^\nu]\Psi$, for $\bar{\Psi}=\Psi^\dagger\gamma^0$ with the gamma matrices $\gamma^\mu$ defined as ${\GAMMA}=\beta{\ALPHA}$, $\gamma^0=\beta$ and $\gamma^5=i\gamma^0\gamma^1\gamma^2\gamma^3$. 
A dimensional analysis shows that the four-components Dirac field has dimensions $[\Psi]=(\text{Length})^{-3/2}$ compatible with the units of the Hamiltonian density above. This fixes the dimensions of the current-current interaction terms (to $(\text{Length})^{-6}$) which, in fact, implies that $[U]=\text{Energy}\cdot(\text{Length})^{3}$. The Hamiltonian in Eq. (\ref{eq:Thirring}) is the tensorial generalization of the Thirring model~\cite{Thirring} in $3+1$ dimensions. This generalization of the Thirring model is not renormalizable (at least by means of perturbative methods). It is analogous in spirit to the  Nambu-Jona-Lasinio model \cite{Nambu1}  and the Fermi effective model \cite{Fermi1} for weak interactions which involves the non renormalizable point-like interaction between two currents. 
The energy scale associated with the Thirring coupling $U$  is given by $\tilde{U}=\sqrt{\frac{(\hbar c)^3}{U}}$ (which is the only energy scale we can define from $\hbar$, $c$ and $U$). This gives a dimensional analysis justification to the  adiabaticity condition $\tilde{U}=\sqrt{\frac{(\hbar c)^3}{U}}\ll m_- c^2$ given above which allows us to restrict the physics around one Fermi point. From now on we will use units where $c=\hbar=1$.\\
\subsection{Microscopic prescription}
\label{section:microscopicPrescription}
We now want to find the microscopic description for the Hamiltonian in Eq. (\ref{eq:Thirring}). To achieve this, we proceed backwards and substitute in $J^\mu$ and $J^{\mu\nu}$ the expressions for the spinor $\Psi=\left(\begin{array}{cccc}a_{{\bf p}} &d_{{\bf p}} &c_{{\bf p}}&b_{{\bf p}}\end{array}\right)^T$ and the gamma matrices as given by $\gamma_0=\beta$ and $\gamma_i=\beta\alpha_i$. After some  tedious calcuations one gets
\begin{equation}
\begin{array}{lll}
J_\mu J^\mu&=&(\bar{\Psi}\gamma_\mu\Psi)(\bar{\Psi}\gamma^\mu\Psi)\\
&=&-2(a^\dagger a+b^\dagger b+c^\dagger c+d^\dagger d)\\
&&+2 (a^\dagger a)(d^\dagger d)+4 (a^\dagger a)(c^\dagger c)+6 (c^\dagger c)(d^\dagger d)\\
&&+6 (a^\dagger a)(b^\dagger b)+4 (b^\dagger b)(d^\dagger d)+2 (c^\dagger c)(b^\dagger b)\\
&&+6a^\dagger b^\dagger  c d+6 c^\dagger d^\dagger a b-2 a d b^\dagger c^\dagger-2 b c a^\dagger d^\dagger\;\;,
\end{array}
\end{equation}
and
\begin{equation}
\begin{array}{lll}
J_{\mu\nu} J^{\mu\nu}&=&(\bar{\Psi}\gamma^5[\gamma_\mu,\gamma_\nu]\Psi)(\bar{\Psi}\gamma^5[\gamma^\mu,\gamma^\nu]\Psi)\\
&=&48\left[(a^\dagger a)(b^\dagger b)+(c^\dagger c)(d^\dagger d)\right.\\
&&-\left.(a^\dagger a)(d^\dagger d)-(b^\dagger b)(c^\dagger c)\right]\\
&&-48\left[a^\dagger b^\dagger c d+b^\dagger c^\dagger a d+c^\dagger d^\dagger a b+ a^\dagger d^\dagger b c\right]\;\;.
\end{array}
\end{equation}
The terms involving only two fermions can be omitted since they give a contribution to the Hamiltonian kernel which is proportional to the identity and can be seen as a constant chemical potential on every site of the lattice. The other terms either interactions between two sites populations or between four sites.\\
This shows the explicit form of the microscopic interaction needed  to simulate the Tensorial Thirring model at low energies given by Eq. (\ref{eq:Thirring}). We note that some of these interactions are attractive while other are repulsive.
\subsection{Bosonization}
\label{section:bosonization}
Throughout the rest of the section we assume that $g\neq 0$ in Eq. (\ref{eq:Thirring}).
The Thirring model describes  relativistic fermions with  self-interactions. In order to get a more accessible theory it is possible to linearize the interaction by introducing new degrees of freedom. Following this approach,  we now show how to describe the low energy physics of the Tensorial Thirring model with a pure bosonic theory. \\
As it can be seen from Eq. (\ref{eq:Thirring}), the  effective theory of our model  is described, in Euclidean space,  by the action
\begin{equation}
\label{eq:Zthirring}
Z_{\text{TTh}}=\int \mathcal{D}[\bar{\Psi}]\mathcal{D}[\Psi]e^{-\mathcal{S}_D-\mathcal{S}_J}\;\;,
\end{equation} 
where the Dirac action, $\mathcal{S}_D=\int d^4x \bar{\Psi}(\slashed\partial-m)\Psi$, and the action for the currents, $\mathcal{S}_{J}=\int d^4x\frac{g^2}{2 m}(2J_{\mu\nu}J^{\mu\nu}-J_\mu J^\mu)$, are given in Euclidean space.  Clearly, $\mathcal{S}_{J}$ involves products of four spinors. To analytically treat this model we linearise the action in terms of the currents by introducing the Hubbard-Stratonovich transformation~\cite{Cantcheff1}. Indeed, we employ the bosonic degrees of freedom $\mathcal{F}\equiv (a^\mu, b^{\mu\nu})$ (in terms of a $4-$vector field  $a^\mu$ and an antisymmetric tensor' field $b^{\mu\nu}$) to write
\begin{equation}
e^{-S_{J}}=\int  \mathcal{D}[\mathcal{F}] e^{\int d^4x\frac{1}{2}(\frac{1}{2}b_{\mu\nu}b^{\mu\nu}-a_\mu a^\mu)+\frac{g}{\sqrt{m}}(b_{\mu\nu}J^{\mu\nu}-a_\mu J^\mu)}\;\;.
\end{equation}
%
%
%
%
Following  \cite{Cantcheff1} we can integrate out the Dirac fermions to find an effective bosonic theory
\begin{equation}
\label{eq:spin1}
Z_{\text{TTh}}=\int D [a] D[ b]e^{-S_{\text{eff}}+\frac{1}{2}\int d^4 x (\frac{1}{2}b_{\mu\nu}b^{\mu\nu}-a_\mu a^\mu)}\;\;,
\end{equation}
where the effective action is defined as
\begin{equation}
S_{\text{eff}}[a_\mu,b_{\mu\nu}]=-\log{\det{(\slashed\partial-m+\frac{g}{\sqrt{m}}\slashed{\mathcal{F}})}}\;\;,
\end{equation}
where  $\slashed{\partial}=\gamma^\mu\partial_\mu$ and $\slashed{\mathcal{F}}=\gamma^\mu a_\mu+ \gamma^5[\gamma^\mu,\gamma^\nu]b_{\mu\nu}$.
Up to terms of order $\partial/m$ ($p/m$ in momentum space) \cite{Cantcheff1}  this effective action can be written as
\begin{equation}
S_{\text{eff}}[a_\mu,b_{\mu\nu}]=-8\frac{g^2}{(4\pi)^2}\int d^4 x \epsilon^{\mu\nu\lambda\sigma}a_\mu\partial_\nu b_{\lambda\sigma}\;\;,
\end{equation}
where $\epsilon^{\mu\nu\lambda\sigma}$ is the Levi-Civita symbol. The correction terms are insignificant in the large-wavelength/low-energy regime we are interested in. By neglecting the irrelevant constants, the partition function for the final theory can  be written as
\begin{equation}
\label{eq:Ztilde}
\tilde{Z}=\int \mathcal{D}[\mathcal{F}]e^{-\int d^4 x \frac{1}{2}(a_\mu a^\mu-\frac{1}{2}b_{\mu\nu}b^{\mu\nu})+\frac{g^2}{2}  \epsilon^{\mu\nu\rho\alpha}a_\mu\partial_\nu b_{\rho\alpha}}\;\;.
\end{equation}
We  note that the fields  $a$ and $b$ have dimension $(\text{Length})^{-2}$. We could be tempted to consider the field $a$ as a sort of electromagnetic field and $b$ as a sort of curvature field. Unfortunately, such an interpretation is not obvious at this stage. In fact, the theory is not invariant under the gauge-like transformation  \cite{Blau1}
\begin{equation}
\left\{\begin{array}{lll}
a_\mu&\rightarrow& a_\mu+\partial_\mu \chi\\
b_{\mu\nu}&\rightarrow& b_{\mu\nu}+\partial_\mu \xi_\nu-\partial_\nu \xi_\mu \;\;,
\end{array}\right.
\end{equation}
where $\chi$ and $\xi_\mu$ are a scalar  and a vector field  respectively. In fact, the kinetic terms $a_\mu a^\mu$ and $b_{\mu\nu}b^{\mu\nu}$ explicitly break invariance under these transformations (as, for example, the vector potential appears explicitely). Hence, the partition function $\tilde{Z}$ describes a massive spin-1 theory that does not allow easy interpretations. We would like to recast this theory in a more suitable form given in terms of a ``vector potential'' and a ``curvature'' field, which naturally leads to the next sections' topic. This process is analogous to the (2+1)-dimensional one where a duality between a self-dual free massive field theory and a topologically massive theory \cite{Deser} has been demonstrated.   \\
As a final note, it is important to stress that it is not possible to apply the bosonization procedure proposed here to free Dirac fermions, i.e. without the presence of the current-current interactions  in Eq. (\ref{eq:Thirring}). This means that we cannot tune the parameter $g^2$ to zero without encountering non-analytical points, which justifies the presence of the factor ${1}/{g^2}$ in Eq. (\ref{eq:Cremmer}). Physically,   the naive replacement $g=0$ into the initial (Eq. (\ref{eq:Thirring})) and final (Eq. (\ref{eq:Ztilde})) theories would lead to a mapping between \emph{free} fermionic degrees of freedom and  \emph{free} bosonic ones, which is clearly forbidden by the statistics of the fields involved. The presented ``transmutation'' of degrees of freedom holds only for interacting theories ($g\neq 0$), as for example happens in superconductivity where the  interaction between electrons in a metal leads to a physics described by bosonic degrees of freedom in terms of Cooper pairs \cite{Cooper1}.
\subsection{Duality}
\label{section:duality}
 In order to recast the theory defined in Eq. (\ref{eq:Ztilde}) in  a more suitable form, one can employ a BFT quantization procedure~\cite{Batalin1,Batalin2} to show the equivalence of the  massive spin-1 theory
\begin{equation}
L=-\frac{1}{4}b_{\mu\nu}b^{\mu\nu}+\frac{1}{2}a_\mu a^\mu + \frac{g^2}{2}\epsilon_{\mu\nu\lambda\sigma} b^{\mu\nu}\partial^{\lambda}a^\sigma\;\;,
\end{equation}
 to one involving an ``electromagnetic'' field $A^\mu$ and a so called Kalb-Ramond field $B^{\mu\nu}$ \cite{KalbRamond}.  The two theories can in fact be embedded in the same enlarged theory from which they descend as different choices of gauge fixing~\cite{Harikumar,KimKim}. The resulting theory, in the Lorentzian signature, is described by the Cremmer-Scherk Lagrangian~\cite{CremmerScherk}
\begin{equation}
\label{eq:Cremmer}
\mathcal{L}=-\frac{1}{4}F_{\mu\nu}F^{\mu\nu}+\frac{1}{12}H_{\mu\nu\lambda}H^{\mu\nu\lambda}+\frac{1}{4 g^2}\epsilon_{\mu\nu\lambda\sigma}B^{\mu\nu}F^{\lambda\sigma}\;\;,
\end{equation}
where $F_{\mu\nu}=\partial_\mu A_\nu-\partial_\nu A_{\mu}$ and $H_{\mu\nu\lambda}=\partial_\mu B_{\nu\lambda}+\partial_\nu B_{\lambda\mu}+\partial_\lambda B_{\mu\nu}$. 
The field $A$ is an effective electromagnetic field (with dimension $(\text{Length})^{-1}$) while the field $B$ (with dimension $(\text{Length})^{-1}$) is the so called Kalb-Ramond field.  As above,  we can define the symmetry transformation
\begin{equation}
\label{eq:gauge}
\left\{\begin{array}{lll}
A_\mu&\rightarrow& A_\mu+\partial_\mu\chi\\
B_{\mu\nu}&\rightarrow& B_{\mu\nu}+\partial_{\mu}\xi_\nu-\partial_{\nu}\xi_\mu\;\;.
\end{array}\right.
\end{equation}
 Contrary to the theory described in Eq. (\ref{eq:Ztilde}), this one is explicitly invariant under this ``gauge'' transformation.\\
Let us now take some time to analyze the terms appearing in the action in Eq. (\ref{eq:Cremmer}). The first two kinematic terms are geometric (the metric appears explicitly) while the last one has a topological nature and it is the standard BF term. This theory is a topological massive gauge theory in $3+1$ dimensions \cite{CremmerScherk,Allen} and represents the natural abelian generalization of Chern Simons-Maxwell theory in $2+1$ dimensions as the Chern Simons theory cannot exist in $3+1$ dimensions. This way of generating mass for the electromagnetic field (through a topological interaction) is an alternative to the Higgs mechanism and can in fact be connected to superconducting phenomena~\cite{Hansson,Balachandran}. The theory is renormalizable \cite{Allen} and explicitely gauge invariant in the bulk. 
\subsection{Bosonization Rules}

\label{section:BosonizationRules}
The possibility to map the Tensorial Thirring model to a massive gauge theory does not come as a surprise. In fact, in the $(1+1)$-dimensional case the massive Thirring model is equivalent to the sine-Gordon massive scalar theory~\cite{Coleman1}, while in $(2+1)$ dimensions is equivalent to the Maxwell-Chern-Simons theory~\cite{Fradkin1}, where, the Maxwell field acquire a mass through a topological mechanism. Motivated by the analogies with the low-dimensional cases we propose the natural generalization of the {\it bosonization rules} to the three-dimensional case. These rules connect the degrees of freedom of the equivalent fermionic and bosonic theories (up to multiplicative factors) in the following way
\begin{table}[h!]
\centering
\begin{tabular}{ccc}
Dimensions&Theory&Bosonization Rules\\
\hline
&&\\
1+1& sine-Gordon& $J^\mu\rightarrow\epsilon^{\mu\nu}\partial_\nu\phi$\\
2+1 & Maxwell-CS&$J^\mu\rightarrow\epsilon^{\mu\nu\lambda}\partial_\nu A_\lambda$ \\
3+1 & Cremmer-Scherk&$\left\{\begin{array}{lll}J^\mu &\rightarrow &\epsilon^{\mu\nu\lambda\gamma}\partial_\nu B_{\lambda\gamma}\\
 J^{\mu\nu} &\rightarrow &\epsilon^{\mu\nu\lambda\gamma}\partial_\lambda A_{\gamma}
 \end{array}\right.$\\
\hline
\end{tabular}
\end{table}\\
Let us take a little more time to emphasize the analogies with the lower dimensional cases and get some  more intuitions on the bosonization procedure. We can note that the Thirring model is always equivalent to some massive theory. In the (1+1)-dimensional case the equivalent theory is a  sine-Gordon massive scalar theory. The equivalence with the Thirring model has been shown by Coleman \cite{Coleman1} (see also \cite{Delpine1} for extension of the proof to the finite temperature case). In (2+1) dimensions the Thirring model has been proven by Fradkin and Schaposnik to be equivalent to a Maxwell-Chern-Simons theory \cite{Fradkin1}. In this last case, the proof relies on a dualization procedure first showed by Deser and Jackiw \cite{Deser} and the equivalent theory corresponds to a massive gauge theory where the mass of the photon is given thanks to the interaction with a topological Chern-Simons term (which, as already mentioned, is an alternative to the Higgs procedure to give mass to a gauge theory). In 3+1 dimensions the Tensorial Thirring model we are studying is going to be equivalent to a so called Cremmer-Scherk model \cite{CremmerScherk} where a Maxwell theory is coupled to a Kalb-Ramond field \cite{KalbRamond} thanks to a BF term. This is, in analogy with the (2+1)-dimensional case, a massive gauge theory where the mass comes from the topological interaction with the Kalb-Ramond field. Note that, in this case, we need two  fields since we are considering two different types of current-current interactions in the fermionic model. These two fields are necessary to generate mass for the gauge theory in a topological fashion \footnote{Without them, it would be impossible to generate such a mass with just one field. In the abelian case this is very easily seen. In fact, with the presence of  one field the only possibilities for a gauge invariant 4-form are $F\wedge F$ and $F\wedge{}^*F$. The first term has a topological nature while the second term is geometric. The topological term can be rewritten as $F\wedge F=d(A\wedge d A)$. This allows to conclude that the addition of such a differential to the lagrangian would not change the equation of motion (since it adds a total derivative to the Lagrangian) and for this reason it cannot possibly add a mass to the gauge field.}. We also note that similar bozonization rules in $d+1$ dimensions were proposed in \cite{Schaposnik1}.
\subsection{Pure Topological Regime}
\label{section:pureTopologicalRegime}
 From now on we work in a regime where the contribution of the topological  BF term in (\ref{eq:Cremmer}) is dominant, i.e. we want to work with energy scales much smaller than $1/g^2$. Intuitively, this suggests that the Maxwell and Kalb-Ramond field have small kinetic energy compared to their (topologically) acquired mass. In such a regime we are sufficiently close to the ground state and the important contributions to the effective theory come from the topological BF term 
\begin{equation}
\label{eq:actionBF}
S_{\text{BF}}=\frac{1}{4g^2}\int_{\mathcal{M}} d^4x\,\, \epsilon_{\mu\nu\lambda\sigma}B^{\mu\nu}F^{\lambda\sigma}\;\;,
\end{equation}
 where $\mathcal{M}$ is the spacetime manifold associated with our theory.
 
\subsubsection{ Boundary Behaviour}
\label{section:boundaryBehaviour}
We now consider the behaviour of our lattice model at its physical boundary. Several approaches are possible based, for example, on the Symanzik method \cite{Amoretti} or on gauge invariance analysis \cite{Momen,Moore}. Focusing on the latter at the bosonic level a BF theory defined on a non-compact space, $\mathcal{M}$, is not manifestly gauge invariant due to contributions from the boundary, $\partial\mathcal{M}$. Restoring gauge invariance generates a $(2+1)$-dimensional BF theory on the boundary, while leaving the bulk theory unchanged~\cite{Momen,Moore}. Here, we show that the bosonization rules allow to infer exactly the same theory on the boundary of the $(3+1)$-dimensional fermionic lattice model (for more detail we refer to Appendix \ref{section:Boundary}). \\
We start by introducing a minimal coupling between the tight-binding fermions and a pure gauge $U(1)$ field $ A^\phi_\mu = \partial_\mu \phi$ parameterised by $\phi$. This coupling extends (\ref{eq:actionBF}), in the continuum limit, by
\begin{equation}
S_{\phi}=\int_{\mathcal{M}} d^4x \,\,J^\mu\partial_\mu \phi \;\;,
 \end{equation}
but it leaves the physics of the model unchanged. We can now employ the bosonization rule $J^\mu\rightarrow \epsilon^{\mu\nu\lambda\gamma}\partial_\nu B_{\lambda\gamma}$, together with Stokes' theorem and an integration by parts to show that
\begin{equation}
S_{\varphi}=\int_{\mathcal{\partial M}} d^3x \,\,\phi\;\epsilon_{\mu\nu\lambda}\partial^\mu B^{\lambda\nu}\;\;,
 \end{equation}
 where here (and throughout the rest of the paper for integrations on the boundary) the indices run through the coordinates that parameterise $\partial\mathcal{M}$.
The field $\phi$ can now be interpreted as a Lagrange multiplier enforcing the condition $dB=0$ on $\partial \mathcal{M}$. This implies that, locally, $B_{\mu\nu}=\partial_\mu\eta_\nu-\partial_\nu\eta_\mu$, which conveniently implies $S_{\varphi}=0$. This means that the possibility to add $S_{\varphi}$ to our action is equivalent to the constraint $dB=0$ on $\partial\mathcal{M}$. We are now ready to find the effective action on the boundary. In fact, we can rewrite the right hand-side of Eq. (\ref{eq:actionBF}) as
\begin{equation}
\frac{1}{4g^2}\left(\int_{\partial\mathcal{M}} d^3x\,\, \epsilon_{\mu\nu\lambda}B^{\mu\nu}A^{\lambda}-\int_{\mathcal{M}} d^4x\,\, \epsilon_{\mu\nu\lambda}\partial^\mu B^{\nu\lambda}A^{\sigma}\right)\;\;,
\end{equation}
which, by restriction on the boundary, implies the following form for the theory on the boundary
\begin{equation}
\label{eq:BFonBoundary}
S_{\partial\mathcal{M}}=\frac{1}{4 g^2}\int_{\partial\mathcal{M}} d^3x\,\, \epsilon_{\mu\nu\lambda}\,\eta^\mu \partial^\nu A^\lambda\;\;.
\end{equation}
This is indeed a $(2+1)$-dimensional abelian BF theory. It is equivalent to a double Chern-Simons theory that describes time-reversal symmetric physics \cite{PalumboBFGraphene} on the boundary.
\subsubsection{ Physical Observables} 
\label{section:physicalObservables}
We now want to identify physical observables associated with the purely topological part, $S_{\text{BF}}$. Gauge invariant observables of the $(3+1)$-dimensional BF theory are given by expectation values of Wilson {\em surface} operators~\cite{Horowitz,Oda1}, which are a generalization of the $(2+1)$-dimensional Wilson loop operators. These observables 
\begin{equation}
W_B=\langle e^{\frac{i}{g^2}\int_{\partial\Sigma} B}\rangle
\end{equation}
 are defined for any two-dimensional boundary $\partial\Sigma$ of a three-dimensional volume $\Sigma$, where $B$ is the Kalb-Ramond field \cite{KalbRamond}. The corresponding fermionic observables are given by 
\begin{equation}
W_\Psi=\langle e^{iq\int_\Sigma d^3x \,\Psi^\dagger \Psi}\rangle\;\;,
\end{equation}
where $q$ is a generic charge of the (string-like) excitations associated with the field $B$. The correspondence is easily proven by an opportunely manipulation of the Noether charge $Q=\int_\Sigma J^0 d^3 x$ (where $J^0=q\Psi^\dagger\Psi$). The joint use of the the bosonization rule $J^0 =\frac{1}{g^2} \epsilon^{ijk0}\partial_\nu B_{ij}$ (where the constant $g^2$ has been introduced for dimensional reasons) and Stokes' theorem immediately leads to $Q=\frac{1}{g^2}\int_{\partial\Sigma}B$. This proves that $q\int_\Sigma d^3x \, \Psi^\dagger \Psi = {1\over g^2}\int_{\partial\Sigma} B$ (where any proportionality constant implicit in definition of the bosonization rules is absorbed inside the charge $q$) or, in other words
\begin{equation}
\label{eq:relationObservables}
W_B = W_\Psi\;\,.
\end{equation}
For more details of this proof we refer to Appendix \ref{section:Observables1}.
It is a well known fact that $W_B = 1$, identically \cite{Blau}. 
Indeed, one can explicitly confirm (see Appendix \ref{section:Observables2}) that 
\begin{equation}
\frac{1}{g^2}\int_{\partial\Sigma} B= 2\pi n\;,\,\,\,\,n\in\mathbb{Z}\;\;,
\label{eqn:flux}
\end{equation} 
for all permissible configurations of $B$. This implies that the charge $q\int_\Sigma d^3x \,\Psi^\dagger \Psi $ inside a volume $\Sigma$ takes discrete values. While this condition gives, as expected, trivial values for the observable $W_\Psi$ it can be employed to distinguish between trivial (product) states and topologically ordered ones~\cite{Palumbo1}. Indeed, product states correspond to a fixed value of $n$ for a given $\Sigma$, while the highly correlated ones can give different values at each measurement. These values of $n$ are experimentally accessible by measuring fermion populations on the vertices of the tight-binding model that are inside $\Sigma$.
\section{Conclusions}
 In summary, we presented a tight-binding model of spinless fermions that has a variety of behaviours. In the absence of interactions it generalises the methodology employed in the $(2+1)$-dimensional Haldane model to the  $(3+1)$-dimensional case giving a chiral topological insulator. In the presence of interactions it gives rise, in the continuum limit, to the $(3+1)$-dimensional BF theory accompanied by a Maxwell term. Our model can be tuned to be in the topological (BF) or the non-topological (Maxwell) regimes, thus being of relevance to both condensed matter and high energy physics. The versatile method we presented for detecting the topological character of the model can become a powerful diagnostic tool for experimentally probing the topological properties of three-dimensional systems.

{\bf Acknowledgements.--} This work was supported by EPSRC and the ARC via the Centre of Excellence in Engineered Quantum Systems (EQuS), project number CE110001013. We would like to thank G.K. Brennen for useful discussions and support.

\appendix
\section{Discrete Symmetries}
In this appendix we brefly analyze the definitions of the three symmetries used to study the model given in Eq. (\ref{eq:hamiltonian}). Following the main text, throughout this appendix we restrict to translationally invariant spinless fermionic systems.\\

\label{AppDiscreteSymmetries}
\subsection{Time-Reversal}
Time-reversal transformations are associated with the inversion of  time. From a physical point of view we want to address whether the system distinguishes a time direction or not. More precisely, given a certain Hamiltonian $H$ and a solution $\Psi(t)$ of the Shr\"{o}dinger equation $H\Psi(t)=i\partial_t \Psi(t)$ we want to know whether a solution $\Psi^\prime(-t)$ of the equation $H\Psi^\prime(-t)=i\partial_{-t} \Psi^\prime(-t)$ also exists. We can then define the (antiunitary) time-reversal operator $T$ by its action on $\Psi(t)$ as $T\Psi(t)=\Psi^\prime(-t)$. \\
Formally, we define a system to be time-reversal symmetric if such an operator $T$ exists such that $H T\Psi(t)=i\partial_{-t} T\Psi(t)$, where $\Psi(t)$ is known to be a solution of the Shroedinger equation $H\Psi(t)=i\partial_t \Psi(t)$. This condition is equivalent to impose $T^{-1} H T\Psi(t)=T^{-1}i T \partial_{-t} \Psi(t)$ which is satisfied if
\begin{equation}
\label{eq:appT}
\left\{\begin{array}{lll}
T^{-1}i T&=&-i\\
{T^{-1}}H T&=&H\;\;.
\end{array}\right.
\end{equation}
 The first condition tells us that the operator $T$ must be antiunitary while the second can be viewed as a restriction on the Hamiltonian. Given these two conditions, the existence of a solution $\Psi(t)$ of the Shr\"odinger equation $H\Psi(t)=i\partial_t \Psi(t)$ implies that $T^{-1} H T\Psi(t)=T^{-1}i T \partial_{t} \Psi(t)$ which in turn implies $ H T\Psi(t)=i\partial_{-t} \Psi(t)$, that is $T\Psi(t)$ satisfies the Shr\"odinger equation with reversed time.\\
The operator $T$ can be written as $T=T_U K$ where $K$ is the complex conjugation operator and $T_U$ a generic unitary operator. From this, it is easy to show that
\begin{equation}
\left\{\begin{array}{lll}
T T^\dagger=\mathbb{I}\\
T^T=T\;\;,
\end{array}\right.
\end{equation}
since we have $T T^\dagger=T_U K K T_U^\dagger=\mathbb{I}$ and $T^T=K T_U^T=K T_U^T K K=T_U K=T$.\\
In the case of spinless fermions, the operator $T_U$ can be chosen to be the identity, so that $T^{-1} f_{{\bf r}}^\dagger T=f^\dagger_{{\bf r}}$ for every generic fermion operator $f_{{\bf r}}$ labeled by its position ${\bf r}$. The action on the Fourier transformed fermion operator $a_{{\bf p}}$ is easily found to be $T^{-1} a_{{\bf p}} T=\sum_{{\bf r}}T^{-1} e^{i{\bf p}\cdot{\bf r}}T a_{{\bf r}}=\sum_{{\bf r}} e^{-i{\bf p}\cdot{\bf r}} a_{{\bf r}}=a_{-{\bf p}}$. Basically, the time-reversal operator maps a particle with momentum ${\bf p}$ to a particle with momentum $-{\bf p}$. The time-reversal action on the Hamiltonian kernel $\bar{H}({\bf p})$ in momentum space follows from $T^{-1}H T=\sum_{{\bf p}}f^\dagger_{-{\bf p}} \bar{H}^{*}({\bf p})f_{-{\bf p}}$, where ${}^{*}$ denotes the complex conjugation which is introduced accordingly to the first of Eqs. (\ref{eq:appT}). The previous identity shows that time-reversal induces an action $\bar{T}$ on the Hamiltonian kernel given by $\bar{T} \bar{H}({\bf p}) \bar{T}^{\dagger}=\bar{H}^{*}(-{\bf p})$ (with $\bar{T}$ unitary  such that $\bar{T}=\bar{T}^\dagger$). Invariance under time-reversal is then equivalent to the request
\begin{equation}
\bar{H}^{*}(-{\bf p})=\bar{H}({\bf p})\;\;.
\end{equation}
Note that the time-reversal  operator for spinless particles is just complex conjugation so that $T^2=\mathbb{I}$.
\subsection{Particle-hole}
In this subsection we define particle-hole symmetry  \footnote{It is important to notice that the  nomenclature is not uniform in the literature as noted in, for example \cite{Sriluckshmy}. For consistency, we chose to follow the nomenclature used in  reference \cite{Ryu2}, which is in fact different from the one used in \cite{Weinberg}. More precisely, the term ``particle-hole'' used in this article and in \cite{Ryu2} reads ``charge conjugation'' in \cite{Weinberg} and \cite{Sriluckshmy}.} following the analysis given in \cite{Weinberg}.
We start by defining a (unitary) charge conjugation transformation. This transformation does not involve any action on spatial or temporal coordinates. 
Under the action of $C$ the operator that annihilates a particle $f$ transforms in the operator that annihilates an antiparticle ${f^\prime}$ as $C^{-1} f_{{\bf p}} C={f^\prime_{{\bf p}}}$. Incidentally, from this definition and from the linearity of the operator we can derive the action in real space:  $C^{-1}f_{{\bf{r}}}C=C^{-1}\sum_{{\bf p}} e^{i{\bf p}\cdot{\bf{r}}}f_{{\bf p}}C=\sum_{{\bf{r}}} e^{i{\bf p}\cdot{\bf{r}}} {f^{\prime}_{{\bf p}}}=f^{\prime}_{{\bf{r}}}$. In our case we identify the antiparticle with a hole with momentum ${\bf{p}}$ by imposing that  $f^\prime_{{\bf p}}=f^\dagger_{-{\bf p}}$ (the creation of a  hole of momentum ${\bf p}$ is equivalent to destroying a particle of momentum $-{\bf p}$),  to finally get: $C^{-1}f^T_{{\bf p}}C=f^{\dagger}_{-{\bf p}}$, where the transpose operator has been introduced to match the notation used so far where creation (annihilation) operators are accommodated in a row (column) vector.\\
The action on the Hamiltonian kernel is given by :$C^{-1}H C$: = :$C^{-1}\sum_{{\bf p}}f^\dagger_{{\bf p}}\bar{H}({\bf p})f_{{\bf p}}C$: =$\sum_{{\bf p}}$:$f^T_{-{\bf p}}C^{-1}\bar{H}({\bf p})C{f^{\dagger T}_{-{\bf p}}}$:=$-\sum_{{\bf p}}f^\dagger_{-{\bf p}}\bar{H}({\bf p})^T f_{-{\bf p}}$, where $:\;\;:$ indicates the normal ordering operator (which imposes creation operators to be on the left of annihilation ones) and where the minus sign takes into account the fermionic statistics. The above identity shows that charge conjugation induces an action $\bar{C}$ on the Hamiltonian kernel given by $\bar{C}\bar{H}({\bf p})\bar{C}^\dagger=\bar{H}(-{\bf p})^*$  (with $\bar{C}$ unitary and such that $\bar{C}=\bar{C}^\dagger$).\\
 A system is defined to be  particle-hole symmetric if $:C^{-1} HC:=:H:$, which implies 
\begin{equation}
\bar{H}(-{\bf p})^*=-\bar{H}({\bf p})\;\;,
\end{equation}
as one can see by comparing the expression given above for $:C^{-1}HC:$ and the expression for the Hamiltonian in momentum space (and taking into account that the Hamiltonian is hermitian). 
We also note that the charge conjugation operator for spinless particles is just complex conjugation so that $C^2=\mathbb{I}$. This condition, together with the  unitarity one implies
\begin{equation}
\left\{\begin{array}{lll}
CC^\dagger=\mathbb{I}\\
C^\dagger=C\;\;,
\end{array}\right.
\end{equation}
\subsection{Chiral Symmetry}
We define a system to have chiral symmetry if there exist a unitary matrix $\bar{C}_s$ that anticommutes with the Hamiltonian kernel in momentum space \cite{Ryu2, RyuArxiv}
\begin{equation}
\label{eq:appendixDefChiral}
\bar{C}_s \bar{H}({\bf p})=-\bar{H}({\bf p}) \bar{C}_s\;\;,
\end{equation}
and such that $\bar{C}_s^2=\mathbb{I}$. This immediately implies that, for each eigenfunction $\Psi_{{\bf p}}$ with energy $E_{{\bf p}}$ there exist an eigenfunction $\bar{C}_s\Psi_{{\bf p}}$ with energy $-E_{{\bf p}}$, since $\bar{H}({\bf p}) \bar{C}_s\Psi_{{\bf p}}=-\bar{C}_s \bar{H}({\bf p})\Psi_{{\bf p}}=-E_{{\bf p}} \bar{C}_s\Psi_{{\bf p}}$. In the context of our model, chiral symmetry reflects a particular structure of the lattice. In fact, a sufficient condition for the existence of this symmetry is the possibility to colour the lattice such that two vertices of the same colour do not have a common link. This property is known as bi-colourability. In this case it is clear that the Hamiltonian kernel can be written in a block off-diagonal form $\left(\begin{array}{cc}0&\cdot\\\cdot&0\end{array}\right)$ which implies the anticommutation with $\sigma_z$.  \\
Another sufficient condition for the presence of chiral symmetry is the existence of both time-reversal and charge conjugation symmetries \footnote{Following our previous comment, we notice that in a context where our ``particle-hole'' symmetry reads ``charge conjugation'', it can also be the case that ``chiral'' symmetry reads ``particle-hole'' symmetry.}. In this case we can define an (antiunitary) operator $C_s=T C$ whose action on the Hamiltonian kernel is given by (see sections above) $\bar{H}\rightarrow \bar{C}_s \bar{H}\bar{C}_s^\dagger$, where $\bar{C}_s=\bar{T}\cdot\bar{C}$. The existence of both time-reversal and particle-hole symmetries implies that the operator $\bar{C}_s$ anticommutes with the Hamiltonian since $\bar{T} \cdot \bar{C} \bar{H}({\bf p})= -\bar{T} \bar{H}^*(-p)\bar{C}=-\hat{H}({\bf p}) \bar{T}\cdot \bar{C}$. As it combines the action of  the time-reversal and charge conjugation operators, chiral symmetry maps a particle with momentum ${\bf p}$ to a hole with momentum $-{\bf p}$.
\section{Behaviour on the Boundary}
\label{section:Boundary}
In this appendix we study the  details of  how to obtain the effective theory describing the boundary of our material. 
Throughout this appendix we  use the differential forms formalism \cite{Nakahara1}. In this language, the bosonic fields introduced in section \ref{section:bosonization} consist of a $1-$form $A$ and a $2-$form $B$.\\ 
Given their importance in the following derivation, we re-write here the  bosonization rules connecting the fermionic  microscopic degrees of freedom and the bosonic effective ones given in Section \ref{section:BosonizationRules}
\begin{equation}
\label{eq:Bosonization2}
\left\{\begin{array}{lll}
J^\mu&\rightarrow &\frac{1}{g^2}\epsilon^{\mu}_{\;\;\nu\lambda\gamma}\partial^{\nu}B^{\lambda\gamma}\rightarrow\frac{1}{g^2}{}^*dB\\
J^{\mu\nu}&\rightarrow &\frac{1}{g^2}\epsilon^{\mu\nu}_{\;\;\;\;\lambda\gamma}\partial^{\lambda}A^{\gamma}\rightarrow\frac{1}{g^2}{}^*dA\;\;,\\
\end{array}\right.
\end{equation}
where ${}^*$ denotes the Hodge operator \cite{Nakahara1} and where we retain the correct dimensions through the coupling $g^2$.\\
We now introduce an example of a procedure to obtain  the theory on the boundary as proposed in \cite{Momen,Moore}. This approach relies on restoring  gauge invariance  on the boundary. 
We then propose a procedure specific to the model presented here which uses the information contained in Eqs. (\ref{eq:Bosonization2}). The advantage of this procedure is that it does not require any additional physical hypothesis on the system.
\subsection{Example: how to restore gauge invariance on the boundary}
\label{section:BoundaryGeneralRestoring}
We start from the BF theory defined in Eq. (\ref{eq:actionBF})
 \begin{equation}
S_{\text{BF}}=\frac{1}{4g^2}\int_{\mathcal{M}} B\wedge F\;\;.
\end{equation}   
Let us begin by showing that the theory is not  gauge invariance on the boundary. The gauge transformation considered here is the one defined in Eq. (\ref{eq:gauge}) that is
\begin{equation}
\label{eq:gaugeForms}
\left\{\begin{array}{lll}
A&\rightarrow& A+d\chi\\
B&\rightarrow& B+d\xi\;\;,
\end{array}\right.
\end{equation}
where $\chi$ is a scalar function and $\xi$ is a  1-form. It is easy to see that, when we add a boundary $\partial M$ to the manifold $M$, the theory is  invariant under the gauge transformations in Eq. (\ref{eq:gaugeForms}) of $A$ alone (since the only dependence of $A$ is through the gauge invariant quantity $F$). Unfortunately, the theory is not invariant under the  generalised gauge transformations  for  the $B$ term in Eq. (\ref{eq:gaugeForms}). 
In fact, under such a transformation, the action changes as 
\begin{equation}
S\rightarrow S+\Delta S\;\;,
\end{equation}
where
\begin{equation}
\Delta S=\int_M d\xi\wedge F=\int_{\partial M}\xi\wedge F\;\;,
\end{equation}
where in the last step we used integration by parts, Stokes theorem and the abelian Bianchi identity $d F=d^2 A=0$. \\
We now want to modify the orginal action to restore gauge invariance on $\partial M$.\\
Following \cite{Momen} and \cite{Moore} (see also \cite{Balachandran}) we now add a boundary  term $\int_{\partial M} B\wedge A$ to the action so that $S^\prime=S+\int_{\partial M} B\wedge A$. This solves the gauge invariance problem for the field $B$ as easily shown with the following
\begin{equation}
\begin{array}{lll}
\Delta S^\prime&=&\Delta S+\int_{\partial M} d\xi\wedge A\\
&=&\int_{\partial M}\xi\wedge F-\int_{\partial M}\xi\wedge F+\int_{\partial M}d(\xi\wedge A)\\
&=&0\;\;,
\end{array}
\end{equation}
where in the last equality we used the Stokes theorem together with the fact that $\partial\partial M=0$. Note that we now have broken the gauge invariance under the transformation on $A$ as can be easily seen by simple inspection of the additional term $\int_{\partial M} B\wedge A$ which is explicitely dependent on the (gauge) field $A$. In order to restore full gauge invariance, we  introduce a new scalar field $\phi$ with the following transformation properties
\begin{equation}
\phi\rightarrow\phi-\chi\;\;,
\end{equation}
where $\chi$ is the same function appearing in the transformation rule for $A$. We now notice that if we redefine $A\rightarrow A^\prime=A+d\phi=D\phi$ we get  $\Delta A^\prime=d\chi-d\chi=0$, which means that the field $A^\prime$ is gauge invariant. We then can define a final gauge invariant action as
\begin{equation}
S_{\text{tot}}=\int_M B\wedge F^\prime+\int_{\partial M}B\wedge A^\prime\;\;.
\end{equation}
Explicitely, the total action is
\begin{equation}
\begin{array}{lll}
S_\text{tot}&=&\int_M B\wedge F^\prime+\int_{\partial M}B\wedge A^\prime\\
&=&\int_M B\wedge dA^\prime+\int_{\partial M}B\wedge A^\prime\\
&=&\int_M B\wedge (dA+d^2\phi)+\int_{\partial M}B\wedge (A+d\phi)\;\;.
\end{array}
\end{equation}
Notice that we have modified the action only on the boundary and that the additional term  breaks time-reversal symmetry (B is even  for time-reversal since it is a sort of ``electric field'' \cite{BaezElectric}, while A is odd) and it is in fact odd under such symmetry if we impose that $\phi\rightarrow -\phi$ under time-reversal. \\
What is the role of the new field $\phi$ in our theory? This field is actually not a dynamical one.  We can see this by calculating its equation of motion. Let us start by computing $\delta_{\phi} S_\text{tot}$. We have
\begin{equation}
\label{eq:EqMotionPhi}
\begin{array}{lll}
\delta_{\phi} S_\text{tot}&=&\int_{\partial M} B\wedge \delta d \phi\\
&=&\int_{\partial\partial M}d(B\wedge\delta\phi)-\int dB\wedge\delta\phi\\
&=&-\int dB\wedge\delta\phi\;\;.
\end{array}
\end{equation}
The equation of motion for the field $\phi$ is given by $\delta_{\phi} S_\text{tot}=0$ which (from Eq. (\ref{eq:EqMotionPhi})) is fulfilled if $dB=0$ on the boundary $\partial M$. This means that the field $\phi$ is nothing but a Lagrange multiplier enforcing the constraint
\begin{equation}
\label{eq:constraintB}
dB=0\;\;\;\;\text{on}\;\;\partial M\;\;.
\end{equation}
We can now suppose that the boundary for our system is $\partial M=\mathbb{R}\times\Sigma$ where the spatial manifold $\Sigma$ is topologically equivalent to $S^2$. Otherwise stated, our boundary is a sphere embedded in  space. The constraint in Eq. (\ref{eq:constraintB})  tells us that $B$ is a closed 2-form on $\partial M$. Since the second de Rham cohomology class on $\partial M$ is non-trivial, we can conclude that $B=d\eta$   locally on $\partial M$ (i.e. $B$ is a pure gauge there) so that the total action becomes
\begin{equation}
S_{\text{tot}}=\int_M B\wedge F+\int_{\partial M}d\eta\wedge A^\prime\;\;,
\end{equation}
where the boundary term is local on the boundary. This allows to conclude that, locally on $\partial M$, the theory is deribed by a BF theory.
\footnote{Incidentally, we notice that this method of imposing gauge invariance on the boundary is not the only possible one. In fact, one could  change the transformation rules of the $B$ field and impose that (trivially) $B\mapsto B$ on the boundary (that is: $B$ is a true ``curvature'' form on $\partial M$). In this case we can avoid adding  the term $\int B\wedge A$.}
\subsection{BF theory on the boundary}
\label{section:BoundaryBF}
 In the previous example, the existence of a $BF$ theory on the boundary was proved by invoking additional  terms on the boundary (which  involve a new scalar field), justified by the requirement of gauge invariance. In this section we want to closely follow this procedure. Specifically, we still want to add a scalar field in order to impose $dB=0$ on the boundary $\partial M$. The main question we want to address is: can we justify the addition of such a field without imposing gauge invariance? We will find a positive answer as a consequence of the bosonization rules given in Eq. (\ref{eq:Bosonization2}). \\
Let us begin by introducing a pure gauge electromagnetic field $A_{\phi}=d\phi$ in the fermionic tight binding model (we stress that we do not actually require the field $A_\phi$ in the system but we introduce it as a pure gauge only to prove that $dB=0$ on $\partial M$). The interaction can be defined by minimal coupling of the fermionic current with the field $A_\phi$. This implies the addition of a term $J_\mu A_{\phi}^\mu$ to the microscopic action. 
Such a term, manipulated through the bosonization rules Eq. (\ref{eq:Bosonization2}) and Stokes theorem gives
\begin{equation}
\begin{array}{lll}
S_\phi&=&\int_M J_\mu A_{\phi}^\mu\\
&=&\int_M {}^* J \wedge A_{\phi}\\
&=&\int_M dB\wedge d\phi\\
&=&\int_M d(B\wedge d\phi)\\
&=&\int_{\partial M} B\wedge d\phi\;\;.
\end{array}
\end{equation}
We  notice that this  additional term contains a scalar field $\phi$. Compared to the example given in the previous section the introduction of this term is now naturally arising from the minimal coupling of the microscopic theory with a pure gauge degree of freedom. The possibility of this result is given by the bosonization rules present in our analysis. This pure gauge is totally arbitrary and does not change the physics of the model.  We can then treat this field as being a Lagrange multiplier enforcing the condition $dB=0$ on the boundary as explained in the example above. We then have $B=d\eta$ locally on $\partial M$ and we can write  the total action of our theory as
\begin{equation}
S_{\text{tot}}=\int_M B\wedge F+\int_{\partial M}d\eta\wedge d\phi\;\;.
\end{equation}
We can now notice that the second term in this expression (the one coming from the minimal coupling) is actually zero since $\int_{\partial M}d\eta\wedge d\phi=\int_{\partial M}d(\eta\wedge d\phi)=\int_{\partial\partial M}\eta\wedge d\phi=0$ (where we used Stokes theorem and the fact that $\partial\partial M=0$). Basically, the gauge field $\phi$ ``lives'' just enough to impose the constraint $dB=0$ on $\partial M$ before quietly ``dying'' without leaving any trace! The action is then given by
\begin{equation}
\begin{array}{lll}
S_\text{tot}&=&\int_M B\wedge F\\
&=&\int_M B\wedge dA\\
&=&\int_M d(B\wedge A)-\int_MdB\wedge A\\
&=&\int_{\partial M} B\wedge A-\int_MdB\wedge A\;\;.
\end{array}
\end{equation}
Since $dB=0$ on the boundary we have that on $\partial M$ the value of the action is just
\begin{equation}
S_{\partial M}=\int_{\partial M} B\wedge A\;\;,
\end{equation}
and, locally
\begin{equation}
\label{eq:T_boundary}
S_{\partial M}=\int_{\partial M} d\eta\wedge A\;\;.
\end{equation}
This means that (locally) on the boundary $\partial M$ our model is described by a $(2+1)$-dimensional BF theory and proves Eq. (\ref{eq:BFonBoundary}) in the main text. The $(2+1)$-dimensional BF theory  is equivalent to a double Chern-Simons theory that describes time-reversal symmetric physics on the boundary.\\
In summary, we have seen that our model is equivalent to one which has a theory on the boundary with a topological BF term. Notice that, without the term $\int B\wedge A$ (described  in section \ref{section:BoundaryGeneralRestoring}) the introduction of the scalar field done in this section is not enough to restore the gauge invariance on the boundary. More precisely, the theory in Eq. (\ref{eq:T_boundary}) is  invariant for gauge transformations involving the field $A$ alone but not for ones involving also the field $B$.\\
\section{Observables}
\label{section:Observables}
This section has two purposes. The first is to prove Eq. (\ref{eq:relationObservables}) which gives a map between microscopic and effective observables. Such a map is desirable because, on the effective side,  it is possible  \cite{Birmingham,Horowitz} to define observables which witness the topological nature of the BF theory. Eq. (\ref{eq:relationObservables}) gives a way to witness these effects in  a microscopic theory.
 The observables in a BF theory are defined as expectation values of Wilson surface operators
\begin{equation}
\label{eq:WilsonSurfaceOperator}
\langle e^{i\frac{q}{g^2}\int_{\partial \Sigma}B}\rangle\;\;,
\end{equation}
where the  surface $\partial \Sigma$ is defined as the boundary of a generic $3$-dimensional spatial manifold $\Sigma$ and where $q$ is a generic charge of the string-like \footnote{The excitations are string-like and not point-like because the two-form field $B_{\mu\nu}$ couples with tensorial currents $J^{\mu\nu}$ and not to the usual vector currents $J^\mu$ as the  electromagnetic field $A^\mu$ does.} excitations associated with the field $B$. In \cite{Birmingham} it is shown that such an expectation value is equal to 1 for the (3+1)-dimensional case considered here. In fact, it represents the trivial case in which the surface $\partial \Sigma$ does not intersect any loop (which would be defined thanks to the point-like excitations associated with the field $A$)  leading to a null linking number. Explicitly we have
\begin{equation}
\label{eq:TrivialWilson}
\langle e^{i\frac{q}{g^2}\int_{\partial M}B}\rangle=1\;\;.
\end{equation}
This brings us to the second purpose of this section: to explicitely check the validity of Eq. \ref{eq:TrivialWilson} for our specific model, or otherwise stated, to prove Eq.\ref{eqn:flux}.
\subsection{\emph{Effective} Noether charge as a topological number}
\label{section:Observables1}
We begin the proof of Eq. (\ref{eq:relationObservables}) by noticing that the Wilson surface observable in Eq. (\ref{eq:WilsonSurfaceOperator}) can be  written as an \emph{effective} Noether charge. The effective Noether charge can be written as a function of the zeroth component of the Noether fermionic current as
\begin{equation}
Q=\int_\Sigma J^0 d^3 x\;\;.
\end{equation} 
The general  expression for the current is obtained by using the bosonization rule for $J$ in Eq. (\ref{eq:Bosonization2}) as
\begin{equation}
\label{eq:CurrentBosonization}
\begin{array}{lll}
J&=&\frac{1}{g^2}{}^*dB\\
&=&\frac{1}{g^2}\partial_\lambda B_{\mu\nu}{}^*[dx^\lambda \wedge dx^\mu\wedge dx^\nu]\\
&=&\frac{1}{g^2}\partial_\lambda B_{\mu\nu}\epsilon^{\lambda\mu\nu}_{\;\;\;\;\;\;\rho}dx^{\rho}\;\;,
\end{array}
\end{equation}
where we used the definitions
\begin{equation}
\left\{\begin{array}{lll}
B&=&B_{\mu\nu}dx^\mu \wedge dx^\nu\\
dB&=&\partial_\lambda B_{\mu\nu}dx^\lambda dx^\mu\wedge dx^\nu\;\;.
\end{array}\right.
\end{equation}
Eq. (\ref{eq:CurrentBosonization}) directly leads to the expression for the zeroth component of the Noether current
\begin{equation}
\begin{array}{lll}
J_0&=&\frac{1}{g^2}\partial_\lambda B_{\mu\nu}\epsilon^{\lambda\mu\nu}_{\;\;\;\;\;\;0}\\
&=&\frac{1}{g^2}\partial^i B^{jk}\epsilon_{ijk0}\;\;.
\end{array}
\end{equation} 
Any  proportionality constant left implicit in the bosonization rules  can simply be used to rescale the parameter $g^2$.  We now note that this is the same expression in coordinates as the exterior differential in the space dimensions of the form $\varphi_t^* B$ where $\varphi_t^*$ denotes the pull back \cite{Nakahara1} of the form  $B$ in a  constant time slice of spacetime under the map $\varphi: \mathbb{R}^3\rightarrow\mathbb{R}^4$ given by $\{{\bf x}\}\mapsto\{{\bf x},t\}$. We in fact simply have
\begin{equation}
\begin{array}{lll}
d\varphi_t^* B&=&d(\varphi_t^*(B_{\mu\nu}dx^\mu dx^\nu))\\
&=&d(B_{ij}dx^i dx^j)\\
&=&\partial_k B_{ij}dx^i dx^j dx^k\\
&=&\partial_k B_{ij}\epsilon^{kij}dx^1 dx^2 dx^3\\
&=&g^2 J_0 d^3 x\;\;.
\end{array}
\end{equation}
We can now finally write the expression for the effective Noether charge as
\begin{equation}
\begin{array}{lll}
Q&=&\int_\Sigma J^0 d^3 x\\
&=&\frac{1}{g^2}\int_\Sigma d(\varphi_t^*B)\;\;.
\end{array}
\end{equation}
Since $J^0=q\Psi^\dagger \Psi$ for a given charge $q$, we can now use these results to identify observables  for the effective topological theory with fermionic physical observables as
\begin{equation}
\label{eq:ObsWilson}
 e^{\frac{i}{g^2}\int_{\partial \Sigma}B}=e^{i  q\int_{\Sigma}\Psi^\dagger \Psi}\;\;.
\end{equation}
In this way we just proved Eq. (\ref{eq:relationObservables}).
\subsection{Check of quantization of the effective Noether charge}
\label{section:Observables2}
In this subsection we further analyze the left hand side of Eq. (\ref{eq:ObsWilson}) in order to check the validity of Eq. (\ref{eq:TrivialWilson}). For simplicity, we start by rescaling the field $\frac{1}{g^2} B\rightarrow B$. We can  consider the embedding $\varphi_\Sigma:\partial\Sigma\rightarrow\mathbb{R}^3$ of the two  dimensional manifold $\partial\Sigma$ in $\mathbb{R}^3$ and include it in the definition of $Q$. In fact, such an embedding induces a pull-back  map $\varphi_\Sigma^*$ \cite{Nakahara1} which takes differential forms defined in $\mathbb{R}^3$ to differential forms defined in $\partial\Sigma$. We get
\begin{equation}
\begin{array}{lll}
Q&=&\int_{\partial\Sigma}  B_{jk}\;\varphi_\Sigma^*[ dx^j \wedge dx^k]\;\;.
\end{array}
\end{equation}
If we introduce coordinates $\theta^1,\theta^2$ on $\partial \Sigma$ and write the pull-back function in coordinates \cite{Nakahara1}, we find
\begin{equation}
\label{eq:Qintegral}
\begin{array}{lll}
Q&=&\int_{\partial\Sigma}  B_{ij}\;\pp{\varphi^i_\Sigma}{\theta^a}\pp{\varphi^j_\Sigma}{\theta^b}d\theta^a \wedge d\theta^b\\
&=&\int_{\partial\Sigma} \tilde{B}_{ab}d\theta^a \wedge d\theta^b\\
&=&\int_{\partial\Sigma} \tilde{B}\;\;,
\end{array}
\end{equation}
with $i,j=1,2,3$ and  $a,b=1,2$ and where, for simplicity, we defined $\tilde{B}$ as a 2-form living on $\partial \Sigma$ \footnote{Notice that the above integral can be written as a flux $\int {\bf B} \cdot d {\bf S}$ with ${\bf B}_i=\frac{1}{2}\epsilon_{ijk} B_{jk}$, $d{\bf S}_i=\epsilon_{ijk}\;\pp{\varphi^j_\Sigma}{\theta^a}\pp{\varphi^k_\Sigma}{\theta^b}d\theta^a \wedge d\theta^b$.} which is simply defined with  a double pull-back on $B$ as $\tilde{B}=\varphi^{*}_\Sigma\varphi^{*}_t B$.\\
Eq. (\ref{eq:Qintegral}) tells us that we have to compute the surface integral of a two form. In a two-dimensional space every spatial  two-form is always closed, that is $d_{\text{space}}\tilde{B}=0$. In our case  $\partial \Sigma=S^2$ which has non-trivial second de Rham cohomology group. This  allows us to conclude that $B$ is exact only locally exact on $\partial \Sigma$. We now define two patches of the sphere labelled $N$ and $S$ respectively around the north and south pole. We  suppose that the two patches intersect on a closed loop $\gamma$ (let us say the equator). From the previous analysis, we can define $\tilde{B}=d\tilde{A}_N$ and $\tilde{B}=d\tilde{A}_S$ on the two patches and write
\begin{equation}
\begin{array}{lll}
Q&=&\int_N{d \tilde{A}^N}+\int_S{d \tilde{A}^S}\\
&=&\int_N(\partial_1 \tilde{A}^N_2-\partial_2 \tilde{A}^N_1)d\theta^1 \wedge d\theta^2\\
&&+\int_S(\partial_1 \tilde{A}^S_2-\partial_2 \tilde{A}^S_1)d\theta^1\wedge d\theta^2\;\;.
\end{array}
\end{equation}
We can now use Stokes theorem and write
\begin{equation}
\label{eq:ANAS}
Q=\int_\gamma \tilde{A}_N-\tilde{A}_S\;\;,
\end{equation}
where, as defined above, $\gamma$ is the common  line where the surfaces $N$ and $S$ intersect. The origin of the minus sign lies in the fact that $\partial \Sigma$ has no boundary so that $\gamma$ has to be taken with different orientations depending if we are integrating  on $N$ or $S$.\\
How are the two ``potentials'' $\tilde{A}^N$ and $\tilde{A}^S$ related on $\gamma$? We know that, in general, our theory is invariant under the symmetry $B\mapsto B+d\xi$ where $\xi$ is a 2-form, which in coordinates reads: $B_{\mu\nu}\mapsto B_{\mu\nu}+\partial_\mu\xi_\nu+\partial_\nu\xi_\mu$. We now remember, from the analysis given above, that  $\tilde{B}=\varphi^{*}_\Sigma\varphi^{*}_t B$. Since the exterior derivative $d$ commutes with the pullback \cite{Nakahara1} we have that $\tilde{B}\mapsto \tilde{B}+d\tilde{\xi}$, where $\tilde{\xi}=\varphi^{*}_\Sigma\varphi^{*}_t \xi$. This transformation has been studied before \cite{Martellini1}  but its connection with a gauge group is not clear and we will in fact do not suppose any association with a gauge group.  Since $\tilde{B}=d\tilde{A}$ the transformation has to  act on the potentials $\tilde{A}$ as $\tilde{A}\mapsto \tilde{A}+\tilde{\xi}$, or in coordinates $\tilde{A}_\mu\mapsto \tilde{A}_\mu+\tilde{\xi}_\mu$ (where, for clarity, we omitted the labels $N/S$). We now require the field $\tilde{B}$ to be single valued on $\gamma$. This can be imposed by writing the simple looking relation:  $\tilde{B}_N=\tilde{B}_S$ (on $\gamma$) which leads to
\begin{equation}
d\tilde{A}_N=d\tilde{A}_S=d\tilde{A}_N+d\tilde\xi\;\;,
\end{equation}
so that $d\tilde{\xi}=0$ on $\gamma$. This means that $\tilde{\xi}$ is closed and locally exact ($\tilde{\xi}=d\chi$)  on $\gamma$ and also allows to write
\begin{equation}
 {\tilde{A}_N}-{\tilde{A}_S}=d\chi\;\;,
\end{equation}
where $\chi$ is a function defined on $\gamma$ everywhere except for a point. Since we can take a point out of the integral over $\gamma$ without affecting the value of the integral we can write, from Eq. (\ref{eq:ANAS})
\begin{equation}
\begin{array}{lll}
Q&=&\oint_\gamma d\chi\\
&=&\oint(\vec{\nabla}\chi)\cdot d\vec{\gamma}\;\;.
\end{array}
\end{equation}
Unfortunately what was done so far did  not give us any contraints on the value of the discontinuity in $\chi$ around $\gamma$. This is a reflection of the fact that we decided (in all generality) not to associate a gauge group to the transformation properties of the field $B$. Nevertheless, we can still obtain such constraint by invoking the  observable nature of the Wilson surface operators $\langle e^{i\int_{\partial \Sigma} B}\rangle$. 
As such, we do not want these observables to be dependent on some "gauge" choice. In particular, we can always use the arguments given above to show that every transformation of the fields implies: $\langle e^{i\int_{\partial\Sigma} B}\rangle \mapsto\langle  e^{i\int_{\partial \Sigma} B} \rangle e^{i\oint_\gamma d\chi}$. 
Since we do not want the value of the observable to be affected by a (generalised) gauge transformation, we have to impose the condition $\int_\gamma d\chi=2\pi n$ (see also \cite{Szabo11,Bergeron1}) which leads to the final result
\begin{equation}
Q=2\pi n\;\;.
\end{equation}
This ends the proof of Eq. (\ref{eqn:flux}) for our specific model.

\bibstyle{plain}

\end{document}